\documentclass[12pt]{article}
\pdfoutput=1
\usepackage[latin1]{inputenc}

\usepackage{amsmath}
\usepackage{amsfonts}
\usepackage{amssymb}
\usepackage{graphicx}
\usepackage{geometry}
\usepackage{amssymb,epsfig}
\usepackage{hyperref}
\usepackage{subfig}

\makeatletter
\renewcommand\section{\@startsection {section}{1}{\z@}%
                                 {-3.5ex \@plus -1ex \@minus -.2ex}
                                   {2.3ex \@plus.2ex}%
                                   {\normalfont\large\bfseries}}
\renewcommand\subsection{\@startsection{subsection}{2}{\z@}%
                                   {-3.25ex\@plus -1ex \@minus -.2ex}%
                                     {1.5ex \@plus .2ex}%
                                     {\normalfont\bfseries}}
\renewcommand\subsubsection{\@startsection{subsubsection}{3}{\z@}%
                                   {-3.25ex\@plus -1ex \@minus -.2ex}%
                                     {1.5ex \@plus .2ex}%
                                     {\normalfont\itshape}}
\makeatother

\def\pplogo{\vbox{\kern-\headheight\kern -29pt
\halign{##&##\hfil\cr&{\ppnumber}\cr\rule{0pt}{2.5ex}&\ppdate\cr}}}
\makeatletter
\def\ps@firstpage{\ps@empty \def\@oddhead{\hss\pplogo}%
  \let\@evenhead\@oddhead 
}
\def\maketitle{\par
 \begingroup
 \def\thefootnote{\fnsymbol{footnote}}
 \def\@makefnmark{\hbox{$^{\@thefnmark}$\hss}}
 \if@twocolumn
 \twocolumn[\@maketitle]
 \else \newpage
 \global\@topnum\z@ \@maketitle \fi\thispagestyle{firstpage}\@thanks
 \endgroup
 \setcounter{footnote}{0}
 \let\maketitle\relax
 \let\@maketitle\relax
 \gdef\@thanks{}\gdef\@author{}\gdef\@title{}\let\thanks\relax}
\makeatother

\numberwithin{equation}{section}

\newcommand{\be}{\begin{equation}}
\newcommand{\bea}{\begin{eqnarray}}
\newcommand{\ee}{\end{equation}}
\newcommand{\eea}{\end{eqnarray}}
\newcommand\beq{\begin{equation}}
\newcommand\eeq{\end{equation}}

\newcommand{\mc}{\mathcal}

\renewcommand{\t}{\tilde}

\newcommand{\vol}{{\rm vol}}
\newcommand{\It}{{\rm Im} \tau}
\newcommand{\Rt}{{\rm Re} \tau}
\newcommand{\Vf}{V_\text{flux}}
\newcommand{\Vb}{V_\text{brane}}
\newcommand{\ai}{\langle \alpha_i \rangle}
\newcommand{\mK}{{\mathcal K}}
\newcommand{\mV}{{\mathcal V}}

\newcommand{\ba}{\begin{align}}
\newcommand{\ea}{\end{align}}
\newcommand{\bg}{\begin{gather}}
\newcommand{\eg}{\end{gather}}
\newcommand{\bseq}{\begin{subequations}}
\newcommand{\eseq}{\end{subequations}}

\renewcommand{\tanh}{\mathop{\rm th}\nolimits}

\begin{document}

\setcounter{page}0
\def\ppnumber{\vbox{\baselineskip14pt
}}
\def\ppdate{\footnotesize{SLAC-PUB-15462, SU-ITP-13/07}} \date{}

\author{Gonzalo Torroba and Huajia Wang\\
[7mm]
{\normalsize \it Stanford Institute for Theoretical Physics }\\
{\normalsize  \it Department of Physics, Stanford University}\\
{\normalsize \it Stanford, CA 94305, USA}\\
[7mm]
{\normalsize \it Theory Group, SLAC National Accelerator Laboratory}\\
{\normalsize \it Menlo Park, CA 94309, USA}\\
}

\bigskip
\title{\bf Black branes in flux compactifications
\vskip 0.5cm}
\maketitle

\begin{abstract}
We construct charged black branes in type IIA flux compactifications that are dual to $(2+1)$--dimensional field theories at finite density. The internal space is a general Calabi-Yau manifold with fluxes, with internal dimensions much smaller than the AdS radius. Gauge fields descend from the 3-form RR potential evaluated on harmonic forms of the Calabi-Yau, and Kaluza-Klein modes decouple. Black branes are described by a four-dimensional effective field theory
that includes only a few light fields and is valid over a parametrically large range of scales. This effective theory determines the low energy dynamics, stability and thermodynamic properties. Tools from flux compactifications are also used to construct holographic CFTs with no relevant scalar operators, that can lead to
symmetric phases of condensed matter systems stable to very low temperatures. The general formalism is illustrated with simple examples such as toroidal compactifications and manifolds with a single size modulus. We initiate the classification of holographic phases of matter described by flux compactifications,
which include generalized Reissner-Nordstrom branes, nonsupersymmetric $AdS_2 \times \mathbb R^2$ and hyperscaling violating solutions.
\end{abstract}
\bigskip
\newpage

\addtocontents{toc}{\protect\setcounter{tocdepth}{2}}

\tableofcontents

\vskip 1cm

\noindent

\vspace{0.5cm}  \hrule
\def\thefootnote{\arabic{footnote}}
\setcounter{footnote}{0}

\section{Introduction}\label{sec:intro}

Black branes provide a fascinating connection between gravitational physics and strongly coupled quantum field theories at finite density. By the AdS/CFT correspondence~\cite{Maldacena:1997re}, a $(d+1)$--dimensional QFT at finite chemical potential for a conserved charge is dual to a charged $d$--dimensional brane in $AdS_{d+2}$. This has created a new area of research at the interface of condensed matter and high energy physics, with powerful  methods that can be applied to systems of strongly interacting fermions~\cite{Hartnoll:2009sz} . This is especially important given the discovery of new materials, such as high $T_c$ superconductors and heavy fermion metals~\cite{highTc1}, which require going beyond the Fermi liquid paradigm.

A very successful ``bottom-up'' approach has been to apply AdS/CFT to phenomenological models of Einstein gravity plus matter fields at finite chemical potential. Varying the matter content and interactions has revealed a rich set of phenomena and striking connections with condensed matter systems. The next step in this program has been to find string theory realizations of such constructions, and determine which of these phenomena can occur in a consistent theory of gravity. The best understood microscopic AdS/CFT dual pairs arise from Freund-Rubin vacua $AdS_{d+2} \times Y$, with $Y$ a positively curved manifold~\cite{Aharony:1999ti}. These solutions have the important property that the size of $Y$ is of order of the AdS radius~\cite{Freund:1980xh}. The bottom-up models can then be realized as \textit{consistent truncations}, where solutions of the $(d+2)$--dimensional theory can be lifted to the full supergravity theory, keeping only a finite number of fields. In this way, it is possible to obtain new supergravity solutions that include the effects of nonzero chemical potentials or other background fields.

However, the situation is not completely satisfactory, since there is no guarantee that the solutions generated in this way are minima of the full theory.\footnote{except in certain special cases where additional symmetries, such as supersymmetry, ensure stability.} Kaluza-Klein (KK) fields not included in the truncation can develop instabilities, and in general it is extremely hard to establish the perturbative stability of these solutions. Ultimately, the reason is that there is no parametric separation between the AdS scale and the internal radius; there is never a $(d+2)$--dimensional theory at low energies and KK modes cannot be decoupled. In this work we will take a different approach: we will construct black branes in string theory which can be described in terms of a $(d+2)$--dimensional \textit{effective field theory} (EFT), namely a theory with a small number of fields valid up to a UV cutoff  that is parametrically larger than the masses and AdS scale. 

Microscopically, this means that the internal dimensions have to be much smaller than the AdS radius, so we have a compactification as opposed to a truncation. At the level of the $(d+2)$--dimensional theory, an EFT for black branes is rather different from a consistent truncation, in that the small number of light fields in the theory determine all the basic low energy properties, its stability and thermodynamics (at least at the perturbative level). The price to pay is that in general the higher dimensional solution is known only approximately; nevertheless, as we explain below, these approximations have negligibly small effects on the low energy dynamics in perturbatively controlled regimes. In this paper we will accomplish the above goal by constructing black branes in flux compactifications of string theory. Flux compactifications~\cite{Frey:2003tf} provide tools beyond the Freund-Rubin mechanism which can give rise to the desired $R_{AdS} \gg R_{KK}$ hierarchy. Examples of $AdS_4$ and $AdS_5$ vacua with small internal dimensions include~\cite{DeWolfe:2005uu,Polchinski:2009ch}. We will focus on asymptotically $AdS_4$ solutions, dual to $(2+1)$--dimensional field theories.

Our investigation is also motivated by constructing models of symmetric phases of matter that can be stable at very low temperatures. Field theory models of non Fermi liquids often have relevant operators and suffer from symmetry-breaking instabilities as the temperature is lowered. Similar issues are encountered in the gravity side, for instance with superconducting~\cite{Hartnoll:2008vx}  or translation-breaking~\cite{Donos:2011bh} instabilities. While these instabilities can lead to interesting broken phases, it is important to develop tools to stabilize symmetric phases by lifting the dangerous relevant operators. For this, we need to find gravity solutions where all the scalar fields have positive masses. Fortunately, during the last decade there has been important progress in developing mechanisms to lift all the light moduli in string compactifications. The motivation has been to construct string vacua that could give realistic models of cosmology or particle physics, and we will apply these results to stabilize phases of condensed matter systems. Combining fluxes and certain orbifold operations we will exhibit simple flux compactifications that describe holographic QFTs where the only relevant operator is the global current that gives rise to the chemical potential, and all scalar operators are irrelevant. Furthermore, the operators that are charged under the chemical potential --the strongly coupled ``electrons''-- have parametrically large dimensions. This class of theories is then an ideal laboratory for studying interesting IR symmetric phases.

\subsection{Basic setup and structure of the paper}

Before beginning our analysis, let us describe the basic setup. We are mainly interested in $(2+1)$--dimensional QFTs at finite density, so we will focus on flux compactifications that admit $AdS_4 \times Y$ solutions. The internal manifold $Y$ is taken to be a six-dimensional Calabi-Yau (CY) manifold. The main motivation for this is that the low energy theory for CY compactifications is very well understood~\cite{Candelas:1990pi}. Moreover, in the perturbative regime of weak string coupling and large volume, the low energy theory depends only on topological information of the manifold, such as dimensions of cycles and cohomology. The EFT for black branes will then apply very generally to all CYs, making it a powerful and simple tool to analyze the low energy physics.

Now we need to explain how the main ingredients required for black brane solutions --a negative cosmological constant and gauge fields-- are obtained in this setup. The internal space is Ricci-flat, so there is no negative contribution to the potential energy from the internal curvature (unlike the case of Freund-Rubin vacua). The negative energy will come from orientifold planes, and balancing their contribution against fluxes (from color branes) can produce AdS vacua. With the goal of stabilizing all the light scalars, we will work with type IIA string theory, where AdS solutions with the desired properties are already known~\cite{DeWolfe:2005uu}.\footnote{Stabilizing all moduli in type IIB CY compactifications requires nonperturbative effects~\cite{Kachru:2003aw}. It would be interesting to understand these instanton effects in the dual QFT.}

Models based on consistent truncations typically have gauge fields from the isometries of the positively curved internal space $Y$. However, CY manifolds have no isometries, so we need to look for other sources of gauge fields. Already in the early works on CY compactifications it was noticed that evaluating the RR potentials on the nontrivial cohomology forms can give rise to gauge fields; see~\cite{Grimm:2004ua} for a review and references. Decomposing the type IIA RR-potential in terms of a harmonic two-form $\omega$,
\be
C_3 = (A_\mu dx^\mu) \wedge \omega\,,
\ee
gives a 4d gauge field $A_\mu$, whose equation of motion follows from the harmonic condition. Therefore, the nontrivial topology of the CY gives rise to gauge fields in the compactified theory. This is the mechanism that we will use in order to have dual field theories at finite chemical potential. Similar topological charges have been employed recently in~\cite{Denef:2009tp}.

The general framework of type IIA flux compactifications on CY manifolds is presented in \S \ref{sec:typeIIA}, focusing  on the structure and interactions of gauge fields and properties of the field theory duals. Most of this material is review, but applications of CY flux compactifications to holographic systems at finite density have not been considered before, so it is useful to have a self-contained exposition. Concrete examples are discussed in \S \ref{sec:models}, which shows that very simple manifolds such as orbifolds of a six-dimensional torus have the required properties to produce black branes. Here one can understand very explicitly the internal geometry, fluxes, and gauge fields. This section also analyzes models that are dual to QFTs with no relevant scalar operators. CYs with minimal matter content (one size modulus and one gauge field) are already of this type. We also suggest a nonsupersymmetric model that captures some of the basic constraints from string compactifications. This theory has only one scalar field and is a useful toy model for exploring possible IR phases.

The main part of the paper is \S \ref{sec:EFT}, where we explain how to obtain black branes in flux compactifications and present the general effective field theory and its regime of validity. Thermodynamic properties of black branes are analyzed using holographic renormalization. Flux compactifications then provide a UV complete and perturbatively controlled framework for classifying holographic phases of matter; we initiate this analysis in \S\ref{sec:solutions}.  We present generalizations of the AdS Reissner-Nordstrom black hole, nonsupersymmetric $AdS_2 \times \mathbb R^2$ solutions, and branes with hyperscaling violation. A lot of work remains to be done to understand the phase structure of black branes in flux compactifications, and in \S \ref{sec:concl} we present a summary and possible future directions.

\section{Type IIA compactifications, gauge fields and holography}\label{sec:typeIIA}

This first section describes the general supergravity setup in which we will construct 4d black branes. We present a short but hopefully self-contained overview of type IIA flux compactifications on Calabi-Yau manifolds; excellent reviews include~\cite{Frey:2003tf}.
Our focus here will be on the structures needed for black branes: gauge fields and their interactions, the effective field theory description, and properties of the holographic QFT duals.

\subsection{Review of ten-dimensional supergravity and compactification}\label{subsec:10d}

The starting point is the string frame 10d action for type IIA supergravity,
\be
S = \frac{1}{2\kappa_{10}^2} \int \sqrt{-g} \left(e^{-2\phi}\left(R + 4 (\partial \phi)^2 - \frac{1}{2} |H_3|^2 \right) - |\t F_2|^2 - |\t F_4|^2 - F_0^2\right) + S_{CS}+ S_{loc}\,,
\ee
where $S_{CS}$ are the 10d Chern-Simons terms, and $S_{loc}$ denotes the contribution from localized sources (O6 planes in our case). The field strengths $\t F_n = F_n^{bg}+ dC_{n-1}+ \ldots$ have contributions from the RR-potentials $C_{n-1}$,  quantized background fluxes $F_n^{bg}$ (specified below), and some extra terms from mixings with $B_2$. We follow the conventions of~\cite{Grimm:2004ua,DeWolfe:2005uu}.

Consider a compactification $M^{9,1}= M^{3,1} \times Y$, where $Y$ is a Calabi-Yau orientifold. Its cohomology determines the matter content of the 4d theory.
First, the two-dimensional cohomology group splits into even and odd forms under the action of the O6 plane; the basis of harmonic forms is denoted by
\be
\omega_\alpha \;\in\;H^{(1,1)}_+(Y)\;,\;\omega_a \;\in\;H^{(1,1)}_-(Y)\,,
\ee
where $\alpha = 1, \ldots, h^{(1,1)}_+$ and $a = 1, \ldots, h^{(1,1)}_-$.
The dual 4-forms $\t \omega^\alpha$ and $\t \omega^a$ satisfy 
\be\label{eq:dualforms}
\int \omega_\alpha \wedge \t \omega^\beta = \delta_\alpha^\beta\;,\;\int \omega_a \wedge \t \omega^b = \delta_a^b\,.
\ee
An important object will be the triple intersection
\be
\kappa_{ABC} = \int \omega_A \wedge \omega_B \wedge \omega_C\,,
\ee
where $A= (\alpha, a)$. We will see shortly that these harmonic forms give rise to scalar fields (K\"ahler moduli) and gauge fields in the 4d theory.

The other nontrivial cohomology group is $H^{(3)}(Y)$, which encodes the complex structure deformations of the CY. These moduli do not mix with the K\"ahler moduli or the gauge fields, so they will not be turned on in black brane geometries as long as they are stable.
To keep the exposition simple let us assume for now that there are no complex structure deformations, i.e. the CY is rigid; shortly we will explain how to add this sector. For a rigid CY, there are only one even and one odd harmonic 3-forms, which we denote by $\alpha_0\;\in\;H^{(3)}_+(Y)\;,\;\beta^0 \;\in \;H^{(3)}_-(Y)$.

Before introducing the orientifold and fluxes, the theory has $\mc N=2$ supersymmetry and each harmonic 2-form gives rise to an $\mc N=2$ vector-multiplet, namely an $\mc N=1$ vector multiplet plus a chiral multiplet. The orientifold then projects out either the vector or the chiral multiplet. Naively, this would suggest that we cannot have simultaneously a scalar and a gauge field. However, there is a simple way out: all that is needed is to have two-forms of both parities under the orientifold. Then, the orientifold projection will keep the scalars in the odd two-forms and the gauge fields in the even forms. The minimal structure is to have two harmonic forms, one of each parity.

Let us now describe the light fields that will be part of the EFT. The zero modes of the p-forms that are allowed by the orientifold are
\be\label{eq:p-fluct}
B_2  =  b^a(x) \omega_a \;,\;
C_3 = \xi(x) \alpha_0+ A^\alpha(x) \wedge \omega_\alpha\,.
\ee
Here $b^a$ and $\xi$ are axions, while $A^\alpha = A^\alpha_\mu dx^\mu$ are $h^{(1,1)}_+$ gauge fields. Since the internal forms are chosen to be harmonic, these fluctuations satisfy the equations of motion for massless fields.  Additional moduli come from fluctuations of the internal metric. K\"ahler deformations are associated to fluctuations of $(1,1)$ type $ \delta g_{i \bar j} dy^i \wedge dy^{\bar j}$ that can be encoded in the K\"ahler form
\be
J = v^a(x) \omega_a\,.
\ee
The complex structure moduli are fluctuations of type $\delta g_{i j}$ that do not respect the K\"ahler condition; they do not mix with the K\"ahler moduli and their effects are discussed below. More details on the moduli space of CY manifolds may be found in~\cite{Candelas:1990pi}. The last light mode corresponds to the zero mode of the 10d dilaton $\phi(x)$.

The allowed background fluxes are
\be\label{eq:fluxes1}
F_0=m_0\;,\;H_3^{bg} = - p \beta_0\;,\;F_2 = -m^a \omega_a\;,\;F_4 = e_a \t \omega^a\,.
\ee
Since type IIA superstring theory has both electric and magnetic sources, these fluxes need to obey Dirac quantization conditions, which means that the coefficients in (\ref{eq:fluxes1}) are proportional to integers~\cite{DeWolfe:2005uu}. Furthermore, the Gauss law for the O6 charge gives
$m_0 p = - \sqrt{8 \pi^2 \alpha'}$.
This implies that the corresponding integer fluxes for $F_0$ and $H_3$ are order one, while two and four-form fluxes are unconstrained.
We note that fluxes in type IIA deform the topological type of the internal space, making it non-K\"ahler or non-complex~\cite{Frey:2003tf}. However, we will argue in \S \ref{subsec:validity}  that these effects can be neglected in the large flux regime.

\subsection{Four-dimensional effective theory}\label{subsec:4dEFT}

We now compactify over $Y$, assuming a metric ansatz of the form
\be
ds^2 = \frac{e^{2\phi(x)}}{\vol(x)} g_{\mu\nu}(x) dx^\mu dx^\nu + g_{i \bar j}(x,y) dy^i dy^{\bar j}\,,
\ee
where $\vol$ is the 6d internal volume 
\be
\vol = \frac{1}{6} \int J \wedge J \wedge J\,,
\ee 
and the prefactor in front of the 4d metric corresponds to choosing 4d Einstein frame with $\kappa_4^2 = \kappa_{10}^2$ --which we set to one in what follows. The effective low energy theory keeps only the zero modes and background fluxes discussed in \S \ref{subsec:10d}. We also neglect backreaction from fluxes or localized sources via warp factors~\cite{Giddings:2005ff}, nonzero torsion classes that deform the topology type of the internal space, and the internal Kaluza-Klein (KK) modes, an approximation that will be self-consistent over a large range of fluxes. The regime of validity of this effective description will be discussed in detail in \S \ref{subsec:validity}. The effective theory simplifies considerably in this limit: the kinetic terms and gauge kinetic functions are given (approximately) by $\mc N=2$ supersymmetry, and the whole theory depends only on integer fluxes and topological data of the CY.

In this approximation, the 4d theory is $\mc N=1$ supergravity with chiral and vector multiplets, as well as a potential generated by the background fluxes~\cite{Grimm:2004ua},
\be\label{eq:Seff1}
S_\text{eff} = \int d^4 x \sqrt{g} \left(\frac{1}{2}R - K_{I \bar J} g^{\mu\nu} \partial_\mu \phi^I \partial_\nu \bar \phi^{\bar J}   - \Vf - \frac{1}{4}\, \It_{\alpha \beta}\, F^\alpha_{\mu\nu} F^{\beta\,\mu\nu} +\frac{1}{4} \,\Rt_{\alpha \beta}\, F^\alpha_{\mu\nu} \t F^{\beta\,\mu\nu}\right)\,.
\ee
The metric signature is $(-+++)$, and $R$ denotes the Ricci scalar of $g_{\mu\nu}$. The fermionic partners are not included here. Also, we are not showing certain boundary terms (see \S\ref{subsec:holo-thermo}) that need to be added to $S_\text{eff}$ in order to have a well-defined variational problem and for performing holographic renormalization. Let us explain in detail the different contributions to this action. 
 
The complex scalars $\phi^I= (t^a, u)$ are the bosonic components of chiral superfields that combine the axions, dilaton, and metric fluctuation:
\be
t^a \equiv b^a + i v^a \;\;,\;\;
u  \equiv  \xi + \sqrt{2} i e^{-D}\;,\; e^D \equiv \frac{e^\phi}{\sqrt{\vol}}\,.
\ee
These `moduli'  acquire nonzero masses from the background fluxes, but --as we discuss below-- are still much lighter than the KK modes.
The K\"ahler potential is given by
\be
K(\phi, \bar \phi) = -\log (\frac{4}{3} \kappa_{abc} v^a v^b v^c)+ 4 D\,,
\ee
and the kinetic term is calculated as $K_{I \bar J} = \partial_I \partial_{\bar J}K$. 

The flux potential is of the general $\mc N=1$ form,
\be
\Vf= e^K (K^{I \bar J} D_I W D_{\bar J} \bar W - 3 |W|^2)\,,
\ee
with superpotential (neglecting nonzero torsion)
\bea\label{eq:Wflux}
W & = & \int \Omega \wedge H_3 + \int J \wedge F_4 - \frac{1}{2} \int J \wedge J \wedge F_2 - \frac{m_0}{6} \int J \wedge J \wedge J \nonumber\\
&=& - p u + e_a t^a + \frac{1}{2} \kappa_{abc} m^a t^b t^c - \frac{m_0}{6} \kappa_{abc} t^a t^b t^c\,,
\eea
where $D_I W = \partial_I W + (\partial_I K) W$.  In order to calculate the potential, note that the $F_2$ contributions
can be set to zero by the redefinitions
\be\label{eq:shift}
\t e_a \equiv e_a + \frac{1}{2} \frac{\kappa_{abc} m^b m^c}{m_0}\;,\;\;\t u \equiv u -\frac{e_a m^a}{pm_0} - \frac{\kappa_{abc} m^a m^b m^c}{3pm_0^2}\;,\;\;\t t^a \equiv t^a - \frac{m^a}{m_0}\,.
\ee
A few algebraic manipulations then give
\bea\label{eq:Vf1}
\Vf&=&  \frac{3e^{4D}}{\mK} \left\{ \frac{K^{a\bar b}}{4}  \left(\t e_a - \frac{1}{2} m_0 \tilde \mK_a\right)\left(\t e_b - \frac{1}{2} m_0 \tilde \mK_b \right)  + \frac{1}{36}(m_0 \mK)^2  +\left(p \t \xi + \frac{m_0}{6} \t \mK -  \t e_a \t b^a\right)^2  \right.\nonumber\\
&+& \left.  \frac{1}{9} (m_0 \mK)^2 K_{a\bar b} \t b^a \t b^b\right\} 
+ \frac{3}{2} \,p^2 \,\frac{e^{2D}}{\mK} - \sqrt{2}\, |m_0 p| \,e^{3D}\,,
\eea
in terms of the quantities
\bea
&& \mK_{ab} = \kappa_{abc} v^c\;,\;\mK_a = \kappa_{abc}v^b v^c\;,\;\mK = \kappa_{abc} v^a v^b v^c \nonumber\\
&& \t \mK_{ab} = \kappa_{abc} \t b^c\;,\;\t \mK_a = \kappa_{abc}\t b^b \t b^c\;,\;\t \mK = \kappa_{abc} \t b^a \t b^b \t b^c\,. 
\eea
Here $\t b^a$ and $\t \xi$ are the shifted axions introduced in (\ref{eq:shift}). This form makes it clear that the properties around the $AdS_4$ vacua are only sensitive to the combination $\t e_a$, so we could just set the $F_2$ flux to zero. Nevertheless, we will keep $F_2 \neq 0$ because it allows for nonzero expectation values for axions, which have
interesting applications to condensed matter systems with parity breaking~\cite{Hartnoll:2007ai}

A crucial property of the string theory potential in Einstein frame is that $V_\text{flux} \to 0$ in the limit when the internal space decompactifies and the string coupling goes to zero. This limit recovers flat 10d space as a vacuum solution. In particular, we stress that (\ref{eq:Vf1}) has no `hard' cosmological constant, i.e. there is no constant term. These properties will have important effects on the allowed black brane solutions.

Finally we come to the gauge fields. From (\ref{eq:p-fluct}), each gauge field comes from a fluctuation of the 3-form potential $C_3 = A^\alpha \wedge \omega_\alpha$. A D2 brane wrapping the 2-cycle supported on $\omega_\alpha$ is electrically charged under $A^\alpha_\mu$, while
a  D4 brane wrapped on the 4-cycle dual to $\t \omega^\alpha$ carries magnetic charge.  The action for the gauge fields and the relative coefficient between the $F_{\mu\nu} F^{\mu\nu}$ and $F_{\mu\nu} \t F^{\mu\nu}$ terms is fixed by supersymmetry.
The dual field strength is defined by
\be
\t F^{\mu\nu} \equiv \frac{1}{2 \sqrt{-g}} \epsilon_{\mu\nu\rho\sigma} F_{\rho \sigma}\,,
\ee
where the inverse $\sqrt{-g}$ makes the $F \tilde F$ contribution to the action independent of the metric. The gauge kinetic function is linear in the K\"ahler moduli,
\be\label{eq:gauge-kinetic}
\tau_{\alpha \beta} = \kappa_{\alpha \beta a} t^a\,.
\ee 
Therefore, a K\"ahler modulus whose associated (odd) 2-form $\omega_a$ has a nonzero triple intersection $\kappa_{\alpha \beta a}$ with two even forms ($\omega_\alpha$ and $\omega_\beta$) will couple to a gauge field. The vacuum expectation value $\langle v^a \rangle$ determines the gauge coupling; likewise, a nonzero axion $\langle b^a \rangle$ gives rise to a $\theta$-angle. 

In summary, type IIA flux compactifications on CY manifolds give rise to a low energy theory (\ref{eq:Seff1}) containing gravity and chiral and vector supermultiplets with specific kinetic terms and gauge kinetic functions (dictated by supersymmetry), as well as a flux potential for the chiral superfields. The CY needs to have cohomology 2-forms of both parities under the orientifold so that the 4d theory contains both scalars and gauge fields. The theory is fully specified in terms of integer fluxes and topological data of the CY manifold, such as the dimensions of cycles and the triple intersection form.

\subsubsection{Supersymmetric $AdS_4$ vacua}

This class of flux compactifications admits $AdS_4$ vacua that preserve 4 supercharges, each of which is dual to a $(2+1)$--dimensional CFT (described in more detail in \S\ref{subsec:CFTduals}). A black brane that asymptotes to such a vacuum near the boundary is then dual to a CFT at finite charge and/or temperature. Let us describe the properties of these vacua in some detail.

Imposing the F-term conditions $D_u W = D_a W=0$ gives the supersymmetry preserving vacua for the K\"ahler moduli
\be
\langle b^a \rangle = \frac{m^a}{m_0}\;\;,\;\; 3 m_0^2 \kappa_{abc} \langle v^b \rangle \langle v^c \rangle + 10 m_0 e_a + 5 \kappa_{abc} m^b m^c=0\,,
\ee
and for the axio-dilaton
\be
p \langle \xi \rangle = \frac{e_a m^a}{m_0} + \frac{\kappa_{abc} m^a m^b m^c}{3m_0^2}\;,\;  \langle e^{-D} \rangle = - \frac{2 \sqrt{2}}{15} \frac{m_0}{p} \langle \kappa_{abc} v^a v^b v^c \rangle\,.
\ee
Therefore, the 10d string dilaton is stabilized at
\be
 g_s = \langle e^\phi \rangle = \frac{15}{4 \sqrt 3} \left| \frac{p}{m_0} \right| \frac{1}{\sqrt{\langle \kappa_{abc}v^a v^b v^c\rangle}}\,,
\ee
and the cosmological constant and AdS radius become
\be\label{eq:LAdS4}
\frac{3}{L^2} = - V_\text{min} = 3 \langle e^K |W|^2 \rangle = \frac{2025}{64} \frac{p^4}{m_0^2} \frac{1}{\langle  \kappa_{abc}v^a v^b v^c \rangle^3}\,.
\ee

One key result of~\cite{DeWolfe:2005uu} is that 4-form fluxes, which are not constrained by charge conservation, can be taken to be parametrically large to obtain an AdS solution with large radius and small string coupling. Indeed, in the large flux limit
\be\label{eq:largeflux}
e_a \sim N \gg 1\,,
\ee
the K\"ahler moduli expectation values scale as $\langle v^a \rangle \sim N^{1/2}$ and then
\be
\vol \sim N^{3/2}\;,\;\;g_s \sim N^{-3/4}\;,\;\;L^2 \sim N^{9/2}\,.
\ee
These scalings are measured in units of $\kappa_4=1$, the 4d Einstein frame that we have adopted.

The masses of light moduli are generically of order of the AdS scale $1/L^2$,
\be
L^2 m_\text{moduli}^2 \sim \mc O(1)\,.
\ee
On the other hand, the mass scale for the KK modes is
\be
m_{KK}^2 \sim \frac{e^{2D}}{(\kappa_{abc} v^a v^b v^c)^{1/3}}\;\;\Rightarrow\;\;L^2 m_{KK}^2 \sim \mc O(N)\,.
\ee
Therefore, at large $N$ the KK scale is parametrically larger than the AdS scale and the masses of the light moduli. It is consistent to treat the vacuum as four-dimensional, and the EFT is valid up to a cutoff of order $m_{KK}$. We refer to these vacua as $AdS_4 \times (small)$.
This fact will be crucial for our construction of the EFT for black branes. In contrast,
Freund-Rubin solutions have $L^2 m_{KK}^2 \sim 1$. Here the vacuum is not four dimensional; in special cases there exist consistent truncations that can be analyzed in 4d, but there is no effective field theory with a finite number of modes.

\subsubsection{Complex structure moduli}

The discussion so far has ignored complex structure moduli; these are not sourced by gauge fields or K\"ahler moduli so they will be spectators of the brane dynamics. Nevertheless, they do interact with the metric and one has to make sure that they do not lead to instabilities. Let us now describe the properties and spectrum of these modes. 

Complex structure deformations are classified by the cohomology group $H^{(2,1)}(Y)$~\cite{Candelas:1990pi} and give rise to $h^{(2,1)}$ chiral superfields in 4d. The pseudo-scalar components are axions coming from evaluating the 3-form potential $C_3$ on the harmonic 3-forms, and the scalar components descend from metric deformations $\delta g_{ij}$. More details may be found in~\cite{Grimm:2004ua}. It is useful to group these together with the dilaton, leading to $h^{(2,1)}+1$ fields $u_k$. From the term $\int \Omega \wedge H_3$ in (\ref{eq:Wflux}), only the linear combination
\be\label{eq:universalD}
W \supset - p_k u_k
\ee
appears in the superpotential. This plays the role of the ``universal'' dilaton $u$ studied before.

In order to find the masses of the complex structure moduli, we will use the following more general result. Consider a chiral superfield $\varphi$; expanding the supergravity potential $V = e^K(K^{\varphi \bar \varphi} |D_\varphi W|^2 -3|W|^2)$ to quadratic order around the supersymmetric vacuum gives
\be
L = K_{\varphi \bar \varphi } |\partial \varphi|^2 + \frac{e^K}{2} 
\left(\begin{matrix}
\varphi &&  \varphi^*
\end{matrix} \right)
\left(\begin{matrix}
 K^{\varphi \bar \varphi} |\partial_\varphi D_\varphi W|^2-2  K_{\varphi \bar \varphi}|W|^2 && -e^K W^* \partial_\varphi D_\varphi W \\
- W (\partial_\varphi D_\varphi W)^*  && K^{\varphi \bar \varphi} |\partial_\varphi D_\varphi W|^2-2  K_{\varphi \bar \varphi}|W|^2
\end{matrix} \right)
\left( 
\begin{matrix}
\varphi^* \\ \varphi
\end{matrix}
\right)
\ee
The physical masses squared are then
\be\label{eq:mpm}
m_\pm^2 = e^K \left(|K^{\varphi \bar \varphi}\partial_\varphi D_\varphi W|^2 \pm |K^{\varphi \bar \varphi}\partial_\varphi D_\varphi W| |W| -2 |W|^2\right)\,.
\ee
This gives the mass splittings between the real and imaginary parts of the chiral superfield $\varphi$. It can also be derived using the supersymmetry algebra of AdS~\cite{deWit:1999ui}.

Now, let $\varphi$ denote a complex structure modulus corresponding to one of the combinations orthogonal to the universal dilaton (\ref{eq:universalD}). The K\"ahler potential only depends on $\varphi - \varphi^*$; the axion cannot appear because of the shift symmetry. Also, $\varphi$ does not appear in the superpotential and hence 
$\partial_\varphi D_\varphi W =- K_{\varphi \bar \varphi}W$.
Eq.~(\ref{eq:mpm}) then gives
\be\label{eq:complex-masses}
m_+^2=0\;,\;
m_-^2 = - \frac{2}{L^2}\,,
\ee
where the AdS radius is $1/L^2 = e^K |W|^2$. We conclude that the mass spectrum of the $h^{(2,1)}$ complex structure moduli is given by massless axions (as expected), and tachyonic modes from the imaginary parts of $\varphi$. See also~\cite{Conlon:2006tq}.
The tachyonic modes do not imply an instability of $AdS_4$ because they are above the BF bound~\cite{Breitenlohner:1982jf} 
\be
m_{BF}^2 = - \frac{9}{4} \frac{1}{L^2}\,.
\ee
In concrete black brane solutions one has to check that such modes do not lead to instabilities. It is also possible to consider rigid CY manifolds, i.e. without complex structure. For instance, in the toroidal models of \S\ref{subsec:toroidal} the complex structure deformations can be projected out by orbifolding, which has the effect of lifting relevant operators in the dual.

\subsection{Three-dimensional CFT duals}\label{subsec:CFTduals}

The $AdS_4 \times (small)$ flux vacua are dual to $2+1$--dimensional conformal field theories with two supercharges, plus two conformal supercharges at the fixed point. This is the smallest supersymmetry in $2+1$ dimensions, a property that is useful for keeping the number of scalar fields to a minimum and improving the stability of the system once supersymmetry is broken by the chemical potentials. An explicit UV Lagrangian for these theories it not yet known, because the stabilization of the dilaton does not allow to interpolate between large 't Hooft coupling (where the gravity solution is valid) and small coupling.
This is similar to $AdS_5$ solutions dual to field theories with dyonic matter~\cite{Fayyazuddin:1998fb,Polchinski:2009ch}.

Nevertheless, many aspects of these CFTs can be calculated from the gravity side. Some properties of the gauge theory can be revealed by trading the 2- and 4-form fluxes by wrapped D6 and D4 branes, which corresponds to the Coulomb branch of the dual field theory~\cite{Fabinger:2003gp}. This analysis was performed in~\cite{Aharony:2008wz}, and further properties of the duals were analyzed by~\cite{Banks:2006hg}.
The central charge of the CFT is of order
\be
c \sim \frac{L^2}{\kappa_4} \sim (\kappa_{abc} v^a v^b v^c)^3 \sim N^{9/2}\,,
\ee
where we used (\ref{eq:LAdS4}) and the large flux limit (\ref{eq:largeflux}); also, recall that we have set $\kappa_4=1$. The dimension of a single trace operator dual to a bulk scalar of mass $m^2$ is determined by the standard quantization formula
\be\label{eq:Delta}
\Delta = \frac{3}{2} + \sqrt{\frac{9}{4}+ L^2 m^2}\,.
\ee

The K\"ahler moduli have $L^2 m_\text{moduli}^2 \sim 1$, so the CFT has a few dual operators with $\Delta \sim\mc  O(1)$.
From (\ref{eq:complex-masses}), the complex structure moduli are dual to $h^{(2,1)}$ marginal operators ($\Delta_-=3$) and their superconformal partners with $\Delta_+=2$.\footnote{Instanton effects from the gravity side can be used to lift the complex structure axions, introducing nonperturbatively small corrections to these dimensions.} The existence of this universal sector of $2 h^{(2,1)}$ operators may shed more light on the UV structure of such CFTs. As described in \S \ref{sec:models}, it is also possible to have gravity solutions with $m^2>0$ from rigid CYs. This situation is also very interesting because the field theory duals have no relevant spin zero operators. The remaining scalar operators are dual to the tower of KK modes; since $L^2 m_{KK}^2 \sim N$, these have $\Delta \sim \sqrt{N}$. Therefore, the CFTs dual to $AdS_4 \times (small)$ gravity solutions have a gap in the dimension of operators. For this reason, these 3d theories are ideal laboratories to study interesting low energy phases of matter at finite density and chemical potential. The simplest CFTs with this property are the minimal models in two dimensions, but in higher dimensions the conformal symmetry is not powerful enough to show the existence of a gap. Holography provides us with such a tool.

The bulk gauge fields $A^\alpha_\mu$ will provide the charges for black branes. They are dual to
 $h_+^{(1,1)}$ global currents $J^\alpha_\mu$ in the CFT. Nonzero boundary values $A^\alpha_\mu(\infty)$ have the effect of adding the source term 
 \be\label{eq:AJterm}
 S_{CFT}\supset \int d^3x\,\sqrt{-g}\,A^\alpha_\mu(\infty) J^\mu_\alpha
 \ee 
 to the field theory dual. (Here $r \to \infty$ denotes the $AdS_4$ boundary.)
A nonzero time component for the gauge field at the boundary is then dual to a chemical potential for the global symmetry. We will also be interested in turning on bulk magnetic fields that do not vanish at the boundary. A spatial component $A^\alpha(\infty) = B^\alpha x_1 dx_2$ gives an external magnetic field $B^\alpha$ for the dual $U(1)$ symmetry.

The field theory origin of these $h_+^{(1,1)}$ global symmetries is interesting: they are dual to three dimensional $U(1)$ gauge fields $a_\mu^\alpha$,
\be\label{eq:Jdual}
J^\alpha = \star_3 d a^\alpha\,.
\ee
(The global symmetry shifts the dual photon by a constant.) The gravity side has been formulated in terms of the gauge fields from the 3-form potential $C_3 = A^\alpha \wedge \omega_\alpha$, but the theory can also be described in terms of the dual gauge fields
\be
d \t A^\alpha = \star_4 dA_\alpha
\ee
that descend from the 5-form $C_5 = \t A_\alpha \wedge \t \omega^\alpha$. Recall that $\t \omega^\alpha$ is the 4-form dual to $\omega_\alpha$, defined in (\ref{eq:dualforms}). The particles electrically charged under $A^\alpha_\mu$ are D2 branes that wrap the 2-cycle $[\omega_\alpha]$, while D4 branes wrapped on $[\t \omega^\alpha]$ are magnetic monopoles --and viceversa for $\t A^\alpha_\mu$. Given this structure, the $U(1)$ gauge fields $a_\mu^\alpha$ arise as the CFT duals of the bulk fields $\t A^\alpha_\mu$~\cite{Witten:2003ya}. This may be seen for instance by integrating by parts the term (\ref{eq:AJterm}) and using the duality relations. An electric charge density for $J_\mu$ is then dual to a density of magnetic flux (from magnetic monopoles) for $a_\mu$, and a background magnetic field for $J_\mu$ corresponds to a chemical potential for $a_\mu$.

It is important to note that the charged particles are not part of the low energy theory in our regime of interest: at large volume and small string coupling they are very massive. This translates to dual ``electrons'' that are described by operators of very high dimension $\Delta \sim N^2$, estimated by placing the wrapped branes as probes in the geometry. This should be contrasted with gauge fields that arise from isometries of internal positively curved spaces and give rise to light charged fields. 

\section{Simple models with no relevant scalar operators}\label{sec:models}

Our discussion so far has been for general CY manifolds. The first question to ask is of course whether there are examples with the structure that we need, namely with $\mc N=1$ orientifolds that lead to both even and odd cohomology 2-forms with nonzero triple intersection. We would also like to have simple string theory models which exhibit the consistency constraints that need to be imposed on bottom-up approaches, and where the properties of black branes and their holographic duals can be analyzed in detail. 

It turns out that already very simple CY manifolds --orbifolds of tori-- have these properties; these models are nice in that the internal geometry and the origin of gauge fields and moduli are very explicit. Also, they have many scalar fields and can lead to a rich phase diagram. After describing these constructions, we turn our attention to a different class of models: CYs with the smallest number of fields, namely one K\"abler modulus and one gauge field. This is the simplest example that can support black branes, and we will analyze its effective field theory in detail. Finally, \S \ref{subsec:simplified} presents a simple one-field model that satisfies basic constraints imposed by string theory but which does not assume supersymmetry.

\subsection{Toroidal orbifolds}\label{subsec:toroidal}

Toroidal orbifolds $T^6/G$, with $G$ a discrete group, can be seen as singular limits of CYs by blowing down certain cycles. The orbifold is chosen to preserve $\mc N=2$ supersymmetry (see~\cite{Reffert:2006du} for a review and references) and then one has to choose an orientifold involution that respects half of the supersymmetries. The orbifold singularities lead to additional complex structure and K\"ahler moduli from twisted sectors, the ``blow-up modes'', that need to be taken into account as well.

As a concrete example, consider the $T^6/\mathbb Z_4$ orbifold with $\mathbb Z_4$ action
\be
(z_1, z_2, z_3) \to \left(i z_1, i z_2, - z_3 \right)\,,
\ee
where $z_i$ are complex coordinates on $T^6$. This model has been studied in detail in~\cite{Blumenhagen:2002gw}.\footnote{We thank T. Wrase for conversations on such models and for pointing out some of these references.} The orientifold action is $\Omega_p (-1)^F \sigma$, where $\Omega_p$ is the worldsheet parity and $\sigma$ is the involution
\be
(z_1, z_2, z_3) \to (\bar z_1,i \bar z_2, \bar z_3)\,.
\ee
Ignoring the blow-up modes for a moment, this model gives rise to four K\"ahler moduli, one complex structure modulus and one gauge field. The nonzero triple intersections between the K\"ahler moduli are
$\kappa_{123} = -\kappa_{344}=\kappa$, i.e. $\vol= \kappa v_3 (v_1 v_2 - v_4^2/2)$; $\omega_3$ has nonzero triple intersection $-\kappa$ with the even two-form, giving rise to a coupling (\ref{eq:gauge-kinetic}) between $v_3$ and the gauge field.

Recall from \S\ref{subsec:4dEFT} that (absent nonperturbative effects) the complex structure moduli have a universal spectrum that leads to relevant operators of dimension 2 and 3 in the dual CFT. As discussed in the introduction, it is important to develop tools to lift relevant operators in order to favor the stability of symmetric phases at low temperatures. In the context of toroidal compactifications, complex structure deformations can be projected out by appropriate orbifolds.
There are various examples of abelian orbifolds with these properties~\cite{Reffert:2006du}. For instance, $T^6/\mathbb Z_8$ plus a suitable orientifold action~\cite{Blumenhagen:2002gw} has two untwisted K\"abler moduli and one gauge field. The 4d field theory is fixed in terms of the triple intersections $\kappa_{122}=\kappa$ for the odd 2-forms, and $\hat \kappa_{111}= - \kappa$ between one odd and two even forms. The orbifold can be understood directly in terms of discrete projections in the dual QFT~\cite{Kachru:1998ys}, providing a mechanism to project out potentially dangerous relevant operators.

One aspect of these constructions is the existence of additional blow-up modes, which quickly increase the number of fields.\footnote{The $T^6/\mathbb Z_4$ model above has 26 twisted $(1,1)$ forms and 6 twisted $(2,1)$ forms, and the  $T^6/\mathbb Z_8$ model has 21 twisted $(1,1)$ forms and no blow-up modes from the $(2,1)$ forms.} The existence of large numbers of fields is presumably a general property of string compactifications, and it would be useful to develop techniques to deal efficiently with this problem. Nevertheless, the situation for toroidal models is slightly simpler than the generic one because the twisted modes behave as probe scalars in the background produced by the untwisted sector. Indeed, taking the 4-form fluxes for the untwisted sector to be order $N$ and the twisted ones to be of order $\hat N \ll N$, the kinetic and mass mixings between twisted and untwisted fields are suppressed by powers of $\hat N/N \ll 1$. A similar suppression arises at the level of the couplings to gauge fields. We can then construct black branes supported by the untwisted sector only (which, as we just saw, can have a very small number of fields), and then add the blow-up modes in the probe approximation. As long as they do not lead instabilities, the black brane solution will be consistent.

\subsection{Models with one K\"ahler modulus}\label{subsec:oneK}

Having discussed some concrete toroidal orbifolds, we now analyze in detail the simplest possible class of CYs with the properties that we need, namely manifolds with one even and one odd harmonic two-form (with nontrivial triple intersection) and no complex structure deformations. The EFT then contains the dilaton and a K\"ahler modulus interacting with a single gauge field. With this matter content the field theory is fully specified (up to the triple intersection number), so there is no need to have a specific model.\footnote{One can check that there are known manifolds with $h^{(1,1)}=2$~\cite{CYexplorer}. It would be interesting to have detailed examples of CY orientifolds with small Hodge numbers.}

Given this field content, the low energy Lagrangian is fixed to
\bea\label{eq:oneK}
L_\text{kinetic}&=&\frac{3}{4v^2}\partial_{\mu}v\partial^{\mu}v+\frac{3}{4v^2}\partial_{\mu}b\partial^{\mu}b+\frac{e^{2D}}{2}\partial_{\mu}\xi\partial^{\mu}\xi+\partial_{\mu}D\partial^{\mu}D- \frac{1}{4} v \,F_{\mu\nu} F^{\mu\nu} + \frac{1}{4} b \,F_{\mu\nu}  \t F^{\mu\nu}\nonumber\\
\Vf&=& \frac{e^{4D}}{\kappa v^3} \left((\t e_1- \frac{m_0}{2}\kappa \t b^2)^2 v^2 + \frac{m_0^2}{12} \kappa^2 v^6 + 3 (p \t \xi + \frac{m_0}{6}\kappa \t b^3 - \t e_1 \t b)^2 + \frac{m_0^2}{4} \kappa^2 \t b^2 v^4 \right) \nonumber\\
&+& \frac{3}{2} \frac{p^2 e^{2D}}{\kappa v^3} - \sqrt{2} |m_0 p| e^{3D}\,,
\eea
where we used the notation introduced around (\ref{eq:Vf1}), and $e_1$ denotes the (only) 4-form flux. Note that the only dependence on the CY enters through the triple intersection $\kappa$, where $6\vol = \mK = \kappa v^3$. The triple intersection between the even and odd harmonic forms has been set to one by a redefinition of the gauge field.

A supersymmetric minimum requires $\text{sgn}(m_0 \t e_1)=-1$, and $\text{sgn}(m_0 p)=-1$ from tadpole cancellation. The F-term conditions stabilize the volume modulus and the dilaton at
\be\label{eq:AdS4vev}
\langle v \rangle = \sqrt{\frac{10}{3}} \sqrt{- \frac{\t e_1}{\kappa m_0}}\;,\; \langle e^{-D} \rangle = \frac{8}{9} \sqrt{\frac{5}{3}} \,\frac{\kappa m_0}{p}\,\left(- \frac{\t e_1}{\kappa m_0} \right)^{3/2}\,,
\ee
and the axions are stabilized at $\langle \t \xi \rangle = \langle \t b \rangle =0$, namely
\be
\langle \xi \rangle = \frac{e_1 m^1}{p m_0} + \frac{\kappa (m^1)^3}{3 p m_0^2}\;,\;\langle b \rangle = \frac{m^1}{m_0}\,.
\ee
Here $m^1$ is the (only) 2-form flux. These expectation values can also be obtained by extremizing (\ref{eq:oneK}). The gauge coupling and $\theta$ angle in the $AdS_4$ vacuum are then given by
\be
\frac{1}{g^2}= \It =\sqrt{\frac{10}{3}} \sqrt{- \frac{\t e_1}{\kappa m_0}}\;,\;\frac{\theta}{8\pi^2} = \Rt= \frac{m^1}{m_0}\,.
\ee
In particular, in the large flux limit $\t e_1 \sim N \gg 1$ the gauge coupling becomes weak.

The $AdS_4$ radius is then given
\be\label{eq:AdS radius}
L^2=\frac{25600\sqrt{\frac{10}{3}}}{2187} \frac{m_0^2 \kappa^3}{p^4}\left( - \frac{\t e_1}{\kappa m_0} \right)^{9/2}\,,
\ee
and expanding the potential (\ref{eq:oneK}) to quadratic order around the supersymmetric vacuum obtains the physical mass eigenvalues
\be\label{eq:massK}
M^2_\text{k}L^2=(70,18)\;,\;\;M^2_\text{axion}L^2=(88,10)\,.
\ee
These $AdS_4$ vacua on CYs with one even and one odd harmonic 2-forms are dual to $(2+1)$--dimensional CFTs with
superconformal primaries of dimensions $\Delta=5$ and $\Delta=10$, a global current, and a gap $\Delta \sim \sqrt{N} \gg 1$ to the rest of the operators.


\subsection{A class of simplified nonsupersymmetric models}\label{subsec:simplified}

So far we have considered theories with $\mc N=1$ supersymmetry, but it is also important to understand what basic properties would still be valid in the absence of supersymmetry (a situation of more general interest). Let us analyze this question for the simple case of a real scalar field $\phi$ interacting with a gauge field $A_\mu$. Based on the properties of the string theory models discussed before, we impose the following conditions on the EFT:
\begin{itemize}
\item The theory is assumed to arise from a weakly coupled supergravity theory, so that the scalar potential and gauge kinetic function will be given in terms of exponentials of the canonically normalized scalar.\footnote{Strong warping can lead to important deformations of the kinetic term~\cite{Douglas:2007tu}  and hence to functional forms different from exponentials.}
\item There should be no `hard' cosmological constant, with the potential asymptoting to zero for $|\phi| \to \infty$. This limit corresponds to the weakly coupled and decompactification limits of string theory.
\item We require the existence of a stable minimum with negative energy.
\end{itemize}
We will find that these simple constraints already lead to a prediction: the absence of relevant spin zero operators in the dual CFT.

The action is then of the form
\be\label{eq:Ssimplified}
S  = \int  d^4x\sqrt{g} \left(\frac{1}{2} R - \frac{1}{2} (\partial \phi)^2 - \frac{1}{4} e^{\alpha \phi} F_{\mu\nu} F^{\mu\nu}  - \Vf \right)\;\;,\;\; \Vf = - A\, e^{-\alpha \nu_1 \phi} + B \,e^{- \alpha \nu_2  \phi}\,,
\ee
where $0< \nu_1 < \nu_2$ and $A$ and $B$ are positive. The range of the scalar field is taken to be $0< \phi < \infty$. It is possible to add more exponential terms to the potential, but they are not required for producing an AdS minimum and do not affect our conclusions in important ways.

The $AdS_4$ vacuum becomes
\be
\alpha \langle \phi\rangle = \frac{1}{\nu_2 - \nu_1} \log \left(\frac{\nu_2 B}{\nu_1 A} \right)\;\;,\;\;V_\text{min} = - \frac{3}{L^2} = - (\nu_2 - \nu_1) \frac{A}{\nu_2} \left( \frac{\nu_1 A}{\nu_2 B}\right)^{\nu_1/(\nu_2 - \nu_1)}\,.
\ee
The dependence on $A$ and $B$ can be absorbed into the AdS radius by shifting the scalar field to zero expectation value,
$\phi = \langle \phi \rangle + \varphi$.
Then,
\be\label{eq:Vfshift}
\Vf =  \frac{3}{L^2} \left( -\frac{\nu_2}{\nu_2 - \nu_1}e^{- \alpha \nu_1 \varphi } +\frac{\nu_2}{\nu_2 - \nu_1}e^{- \alpha \nu_2 \varphi }\right)\,.
\ee
Finally, from (\ref{eq:Vfshift}) we obtain the mass of the scalar field around the minimum
$L^2 m^2 = 3 \alpha^2 \nu_1 \nu_2 >0$ which, as anticipated, is dual to an irrelevant single-trace operator. The dual CFT has no relevant spin zero operators and a global conserved current. 

\section{The effective field theory for black branes}\label{sec:EFT}

Now we have assembled all the necessary tools to construct black branes, and in this section we describe their 4d effective theory. We also perform the holographic renormalization for these solutions and extract the thermodynamic properties of the dual QFTs at finite temperature, chemical potential and magnetic field.

\subsection{Dyonic black branes}\label{subsec:dyonic}

The starting point is the low energy theory for the light moduli
\be\label{eq:Seff2}
S_\text{eff} = \int d^4 x \sqrt{g} \left(\frac{1}{2}R - K_{I \bar J} g^{\mu\nu} \partial_\mu \phi^I \partial_\nu \bar \phi^{\bar J}   - \Vf - \frac{1}{4}\, \It_{\alpha \beta}\, F^\alpha_{\mu\nu} F^{\beta\,\mu\nu} +\frac{1}{4} \,\Rt_{\alpha \beta}\, F^\alpha_{\mu\nu} \t F^{\beta\,\mu\nu}\right) + S_b\,.
\ee
This is an ``Einstein-Maxwell-dilaton-axion'' theory\footnote{The name `dilaton', used in bottom-up models, refers to a field that is neutral under the gauge symmetry, and should not be confused with the string dilaton.} with specific couplings, kinetic terms and flux potential dictated by supersymmetry, topological data of the internal CY, and integer fluxes.  $S_b$ denotes boundary contributions (to be discussed below) that are required for holographic renormalization. 
Dilatonic branes have been extensively studied in the literature, mostly from a bottom-up perspective~\cite{Taylor:2008tg,Goldstein:2009cv,Kiritsis}. There is by now an impressive list of microscopic and macroscopic models, e.g.~\cite{examples} just to cite a few.

We have explained how (\ref{eq:Seff2}) arises as the low energy limit of 10d supergravity compactified on CY manifolds. This effective theory is valid up to a cutoff of order of the KK scale, which is parametrically larger than the cosmological constant. Above this scale, the UV completion is given by classical 10d supergravity. Moreover, quantum and $\alpha'$ corrections are negligible in the regime of small string coupling and large internal radius where the moduli are stabilized.
We now construct black branes sourced by nonzero electric and magnetic charges for the gauge fields $A^\alpha_\mu$. The gauge kinetic function $\tau$ depends on the light fields $\phi^I$, and this coupling will in turn induce a nontrivial radial dependence on the various scalars. The metric ansatz that describes the brane is
\be\label{eq:metric-ansatz}
ds^2 = - e^{-w(r)} f(r) dt^2 + \frac{dr^2}{f(r)} + \frac{r^2}{L^2} (dx_1^2 + dx_2^2)\,,
\ee
and the solution is required to asymptote to the $AdS_4$ vacua of \S \ref{subsec:4dEFT} at large $r$. Here $L$ is the $AdS_4$ radius.

The equations of motion for the metric and gauge field are
\bea
&& R_{\mu\nu}= {\rm Im} \mc \tau_{\alpha \beta} \left(F_{\mu\rho}^\alpha F_{\nu}^{\beta\,\rho}- \frac{1}{4} g_{\mu\nu} F^\alpha_{\lambda \sigma} F^{\beta\,\lambda \sigma}\right)+ 2 K_{I \bar J} \, \partial_\mu \phi^I \partial_\nu \bar \phi^J + g_{\mu\nu} V_\text{flux}\nonumber\\
&&\partial_\mu \left(\sqrt{g}\,{\rm Im}\,\mc \tau_{\alpha \beta}\, F^{\beta\,\mu\nu} \right) = \frac{1}{2} \epsilon_{\mu\nu\rho\sigma} \partial_\mu \left({\rm Re}\,\mc \tau_{\alpha \beta} F^\beta_{\rho \sigma} \right)\,.
\eea
The gauge field equation allows for the following electric and magnetic charges,
\be\label{eq:charges}
F_{tr}^\alpha =  e^{-w/2} \frac{L^2}{r^2}\,\It^{\alpha \beta} (Q_\beta - \Rt_{\beta \gamma} P^\gamma)\;,\;F_{12}^\alpha = P^\alpha\,,
\ee
where $\It^{\alpha \beta} = (\It ^{-1})_{\alpha \beta}$.
The equations of motion for the scalar fields are
\bea
0&=&\frac{1}{\sqrt g} \partial_\mu (K_{I \bar J} \sqrt{g} g^{\mu\nu} \partial_\nu \phi^I)-(\partial_{\bar J} K_{ A\bar B}) g^{\mu\nu} \partial_\mu \phi^A \partial_\nu \bar \phi^B- \partial_{\bar J} V_\text{flux}\nonumber\\
&&\;\;- \frac{1}{4} \,\partial_{\bar J}({\rm Im}\,\mc \tau_{\alpha \beta})\, F^\alpha_{\mu\nu} F^{\beta\,\mu\nu} +\frac{1}{8\sqrt g}\, \partial_{\bar J}({\rm Re}\,\mc \tau_{\alpha \beta})\, \epsilon_{\mu\nu\rho\sigma}F^\alpha_{\mu\nu} F^\beta_{\rho\sigma}\,.
\eea

Computing the Ricci tensor for (\ref{eq:metric-ansatz}) shows that both $R_{tt}$ and $R_{rr}$ depend on second derivatives of $f$ and $w$, while $R_{ij}$ only depends on first derivatives. Taking a linear combination of the $R_{tt}$ and $R_{rr}$ equations that depends only on first derivatives of $f$ and $w$, and combining with $R_{ij}$, allows to determine these functions directly in terms of first order equations
\bea\label{eq:radialeqs}
&& \frac{\partial_r f}{r}+ f \left(\frac{1}{r^2}+ K_{I \bar J} \partial_r\phi^I \partial_r\bar \phi^J \right)+  \frac{L^4}{r^4}\Vb+\Vf=0\nonumber\\
&& \partial_r w + 2 r K_{I \bar J} \partial_r\phi^I \partial_r\bar \phi^J =0\,,
\eea
where we introduced the potential $\Vb$ induced by the charges (\ref{eq:charges}):
\be
\Vb \equiv \frac{1}{2} \,\It^{\alpha \beta} (Q_\alpha - \Rt_{\alpha \gamma}P^\gamma)(Q_\beta - \Rt_{\beta \delta}P^\delta) + \frac{1}{2}  \It_{\alpha \beta} P^\alpha P^\beta\,.
\ee
These equations can be summarized compactly by the radial Lagrangian
\be\label{eq:radialL}
{\mathcal L} = - \frac{r^2}{L^2} e^{-w/2} \left( \frac{\partial_r f}{r}+ \frac{f}{r^2} + f K_{I \bar J} \partial_r\phi^I \partial_r\bar \phi^J +V_\text{eff}(r,\phi)\right)\,,
\ee
with an effective potential
\be\label{eq:Veff}
V_\text{eff}(r,\phi) \equiv  \frac{L^4}{r^4}\Vb(\phi) + \Vf(\phi)\,.
\ee
The gauge fields have been integrated out exactly using (\ref{eq:charges}).

The Lagrangian (\ref{eq:radialL}) determines the radial profile of the black brane, after supplementing the evolution equations by appropriate boundary condition. We require that for $r \to \infty$ the solution asymptotes to the $AdS_4$ vacuum of \S \ref{subsec:4dEFT}, namely
\be
f(r) \to \frac{r^2}{L^2}\;,\; w(r) \to 0\;,\;\phi^I(r) \to \langle \phi^I \rangle\,.
\ee
In general, the solution will develop a horizon at some infrared value $r=r_h$, where $f(r) \propto r - r_h$. The remaining boundary conditions come from requiring regularity at $r=r_h$. In general it is not possible to find analytic solutions to this system of equations, but approximate and numeric solutions will be studied in \S \ref{sec:solutions}.

\subsection{Holographic renormalization and thermodynamics}\label{subsec:holo-thermo}

Let us now study the thermodynamics of black branes and their QFT duals, using holographic renormalization~\cite{Henningson:1998gx}. The boundary term $S_b$ in (\ref{eq:Seff2}) has two kinds of contributions: the Gibbons-Hawking-York boundary term required to have a well-defined variational problem~\cite{York:1972sj}, plus boundary counterterms that subtract the infinities of the on-shell action in the limit $r \to \infty$. The gauge fields do not lead to divergences, and the same is true for scalars assuming (as we do here) that $\Delta>3/2$ as in the examples before.\footnote{The procedure to subtract scalar field divergences for $\Delta <3/2$ is standard and can be straightforwardly implemented in our context.} The only counterterms are those associated to the gravitational field, 
\be
S_{ct}= \int_{\partial M_{d+1}} \sqrt{\gamma}\, \left(\frac{d-1}{L} + \ldots \right)\,,
\ee
where $\gamma_{\mu\nu}$ is the induced metric on the boundary $\partial M_{d+1}$, and the dots are additional curvature invariants that will not be relevant for us.  The on-shell action is evaluated at a fixed (large) radial cutoff $r_c$, and the limit $r_c \to \infty$ is taken at the end. 
With the GHY and counterterm contributions in place, (\ref{eq:Seff2}) gives a finite (cutoff independent) answer. This is the gravitational version of the renormalization procedure in the QFT side.

The holographic stress tensor is given in terms of the extrinsic curvature $\Theta_{\mu\nu}$ of the boundary~\cite{Brown:1992br}:
\be
T_{\mu\nu}= - \left( \Theta_{\mu\nu}-\Theta \gamma_{\mu\nu} + \frac{2}{\sqrt{\gamma}} \frac{\delta S_{ct}}{\delta \gamma^{\mu\nu}}\right)\,.
\ee
The conserved mass associated to $T_{\mu\nu}$, which is dual to the energy of the QFT, is
\be
E = \int_\Sigma \sqrt{\gamma}\, u^\mu u^\nu T_{\mu\nu}\,,
\ee
where $\Sigma$ is a spacelike surface in $\partial M$, and $u^\mu$ is the timelike unit normal to $\Sigma$.

The metric (\ref{eq:metric-ansatz}) has extrinsic curvature components
\be
\Theta_{tt}=- \frac{1}{2} f^{1/2} \partial_r (e^{-w} f)\;,\;\Theta_{ij} = \delta_{ij} \frac{r}{L^2} f^{1/2}\,.
\ee
This gives an energy density
\be\label{eq:energy1}
\varepsilon=  \frac{E}{\mV} = \frac{2}{L} e^{-w/2} \left(f^{1/2}- \frac{L}{r} f \right) \frac{r^2}{L^2}\,,
\ee
where all the quantities are evaluated at $r \to \infty$ and $\mV$ is the two-dimensional spatial volume. Eq.~(\ref{eq:energy1}) vanishes for the $AdS_4$ solution, so the counterterms effectively subtract the diverging contribution of $AdS_4$. In a nontrivial black brane geometry, the only nonvanishing contribution comes from the subleading term
\be\label{eq:asymp}
w \sim 0\;,\;f \sim \frac{r^2}{L^2} - f_1 \frac{L}{r}
\ee
and gives
\be
\varepsilon=  \frac{f_1}{L}\,.
\ee
The expression (\ref{eq:asymp}) is familiar from the Schwarzschild AdS geometry, in which case $f_1$ gives a nonzero temperature. We will soon determine the thermodynamic parameters in more detail.

Finally, let us evaluate the on-shell action of the gravity solution, which gives the thermodynamic potential. The Ricci scalar for (\ref{eq:metric-ansatz}) is
\be
\sqrt{g} R =- \frac{2}{L^2} e^{-w/2} (r \partial_r f + f) - \frac{1}{L^2} \partial_r \left[r^2 e^{w/2} \partial_r (e^{-w}f) \right]\,.
\ee
The first term here also appeared as the gravitational contribution to the radial Lagrangian (\ref{eq:radialL}), while the second term is total derivative that was dropped at that stage because it does not contribute to the equations of motion. However, when evaluated on-shell, only the total derivatives contribute because the remaining terms vanish by the constraint $\delta L / \delta w=0$. The other contributions come from the boundary terms $S_b$ and a total derivative from the Yang-Mills term.\footnote{There is also a total derivative from the scalar fields, which gives a vanishing contribution given our assumption $\Delta>3/2$.} The gravitational terms combine to give
\be
- \frac{1}{2} \int_{M_4} \sqrt{g} R + \int_{\partial M_4} \sqrt{\gamma} \left(\Theta + \frac{2}{L} \right)=\int d^3 x \left\{\left. 2  e^{-w/2} \frac{r^2}{L^2} \left(\frac{f^{1/2}}{L}- \frac{f}{r} \right) \right|_{r \to \infty} - \left. \frac{1}{2}  \frac{r^2}{L^2} e^{-w/2} f' \right|_{r=r_h}\right\}
\ee
For a thermal circle of length $\beta$ (determined below) and recalling the asymptotic behavior (\ref{eq:asymp}), we arrive to
\be
S_{GR}= \beta \mV \left( \frac{f_1}{L} - \frac{1}{2} \frac{r_h^2}{L^2} e^{-w_h/2} f'_h \right)\,.
\ee

Lastly, the contribution from the total derivative in the Yang-Mills sector is
\be
\frac{1}{4} \int \sqrt{g} \left( \It_{\alpha \beta}\, F^\alpha_{\mu\nu} F^{\beta\,\mu\nu} +i\,\Rt_{\alpha \beta}\, F^\alpha_{\mu\nu} \t F^{\beta\,\mu\nu} \right) =  \int d^3 x A_\mu J^\mu
\ee
with the conserved current
\be\label{eq:conserved-current}
J^\mu = \left.\frac{1}{2} \sqrt{g} \left(\It_{\alpha \beta} F^{\beta\,r\mu}+i \Rt_{\alpha \beta} \t F^{\beta\,r\mu} \right)\right|^\infty_{r_h}\,.
\ee
Note that the nonzero $\theta$ angle from the axion $\Rt_{\alpha \beta}$ gives an additional contribution to the current, proportional to $\t F$. Taking into account the regularity condition $A_t(r_h)=0$ and the on-shell value of the field strength, this evaluates to $\frac{1}{2}\,\beta \mV  A^\alpha_t(\infty) Q_\alpha$.
Putting these contributions together obtains
\be\label{eq:S-onshell}
S_\text{on-shell}= \beta \mV \left( \frac{f_1}{L} - \frac{1}{2} \frac{r_h^2}{L^2} e^{-w_h/2} f'_h- \frac{1}{2} A^\alpha_t Q_\alpha\right)\,.
\ee

The previous results determine the thermodynamic properties of black branes and the holographic dual.  Since we have fixed the asymptotic value of $A_t$, which is dual to the chemical potential, the gravitational action will yield the thermodynamic potential $\Omega$ in the grand canonical ensemble.\footnote{We could also work at fixed charge density; this requires adding a boundary term to the action in order to have a well-defined variational problem. The effect of this boundary contribution is to shift $\Omega$ to the free energy $F = \Omega + \mu^\alpha N_\alpha$.}
As usual, the temperature is obtained by expanding the metric around the horizon and requiring the absence of a conical singularity from the shrinking thermal circle. Doing so obtains
\be\label{eq:temperature}
T = \frac{1}{4\pi}f'(r_h) e^{-w_h/2}\,.
\ee
Next, the entropy is obtained from the area of the horizon,
\be
s = \frac{\mc S}{\mV}= \frac{2\pi}{\kappa_4^2} \frac{r_h^2}{L^2}\,, 
\ee
and we work in units of $\kappa_4^2 =1$.

The gauge symmetries of the bulk theory are dual to global symmetries in the CFT, and the asymptotic values of the gauge fields $A_\mu^\alpha$ determine the chemical potential and external magnetic fields for the $\alpha$-th global symmetry:
\bea\label{eq:chemical potential}
\mu^\alpha&=& A_t^\alpha(\infty) = \int_{r_h}^\infty dr\,e^{-w/2}  \frac{L^2}{r^2}\,\It^{\alpha \beta} (Q_\beta - \Rt_{\beta \gamma} P^\gamma)\,.
\eea
From (\ref{eq:conserved-current}), the conserved charge density is
\be
\rho_\alpha= \langle J^t_\alpha \rangle = \frac{\delta S}{\delta A^\alpha_t(\infty)}= - \frac{Q_\alpha}{2}\,.
\ee
The gauge field also has a nonzero component at infinity $A^\alpha \supset P^\alpha x_1 dx_2$, which implies that there is a nonzero external magnetic field for the corresponding $U(1)$ symmetry,
\be\label{eq:magneticB}
B^\alpha = P^\alpha\,.
\ee

In the grand canonical ensemble, the thermodynamic potential is related to the on-shell action by
\be
\Omega = T S_\text{on-shell}= \mV \left( \frac{f_1}{L} - \frac{1}{2} \frac{r_h^2}{L^2} e^{-w_h/2} f'_h+ \frac{1}{2} A^\alpha_\tau Q_\alpha\right)\,.
\ee
The right hand side of this expression is recognized as the definition of the grand potential
$\Omega = \mV \left(\varepsilon - T s - \mu^\alpha \rho_\alpha \right)$. The pressure is given by
\be
P = - \frac{\Omega}{\mV}= -\left( \frac{f_1}{L} - \frac{1}{2} \frac{r_h^2}{L^2} e^{-w_h/2} f'_h+ \frac{1}{2} A^\alpha_\tau Q_\alpha\right)
\ee
that follows from the familiar relation $\varepsilon= T s - P + \mu^\alpha \rho_\alpha$.

The magnetic field induces a magnetization $M_\alpha = m_\alpha \mV$. Recalling that
\be\label{eq:energy2}
d \varepsilon = T ds + \mu^\alpha d\rho_\alpha - m_\alpha dB^\alpha\,,
\ee
the magnetization density $m_\alpha$ is given by
\be
m_\alpha = -\left. \frac{\partial \varepsilon}{\partial B^\alpha} \right|_{s,\,\rho} = - \frac{1}{L}\left. \frac{\partial f_1}{\partial B^\alpha} \right|_{s,\,\rho}\,.
\ee
Calculating this quantity requires knowledge of the full numerical solution. Nevertheless, to get some intuition, let us assume that $m^2>1/L^2$, and approximate the metric by the RN-AdS warp factor,
\be
f(r) \approx \frac{r^2}{L^2} - f_1 \frac{L}{r} + L^2 \langle \Vb \rangle \frac{L^2}{r^2}\,.
\ee
Evaluating this at $r=r_h$ determines $f_1$ and hence the energy,
\be
\varepsilon \approx \frac{r_h}{L^2}  \left( \frac{r_h^2}{L^2}+ L^2 \langle \Vb \rangle \,\frac{L^2}{r_h^2}\right)\,.
\ee
Since $s \propto r_h^2$, and the charge and magnetic field appear as explicit variables in $\Vb$, we obtain, to linear order in the charges,
\be
m_\alpha = \frac{3}{4\pi T}\left(2\langle \Rt_{\alpha \beta} \It^{\beta \gamma} \rangle \rho_\gamma - \langle \It_{\alpha \beta}- \Rt_{\alpha \gamma} \It^{\gamma \delta} \Rt_{\delta \beta}\rangle B^\beta \right)\,,
\ee
where $\frac{r_h}{L^2} \approx \frac{4\pi}{3}T$ at this order. We thus find a magnetic susceptibility controlled by the gauge coupling and $\theta$ angle, as well as a contribution from the charge density when the $\theta$ angle is nonvanishing.

\subsection{Regime of validity of the effective theory}\label{subsec:validity}

Lastly, let us determine the regime of validity of the black brane EFT. As we discussed before, the 4d action (\ref{eq:Seff2})
arises as the low energy limit of type IIA 10d supergravity compactified on CY manifolds. The UV cutoff is set by the KK scale
$m_{KK}^2 \sim N/L^2$,
where $N$ is the order of magnitude of the 4-form flux, and the fields kept in the effective description have masses $m^2 \sim 1/L^2$. For $N \gg 1$ the UV cutoff is parametrically larger than the AdS scale and light masses, and KK modes can be decoupled. The theory also receives $\alpha'$ and quantum corrections, but these are parametrically small at large $N$.  Another source of corrections comes from the localized sources (e.g. the O6 plane), which backreact on the metric and RR-potentials. The compactified theory solves the equations of motion on average, and the backreaction appears in the form of  warp and conformal factors and nontrivial internal wavefunctions for the dilaton and p-form potentials, which allow to solve the equations of motion pointwise. These effects, which correct the kinetic terms and flux potential, are small at small $g_s$ and large volume, as follows from the general results of~\cite{Giddings:2005ff}.

Furthermore, fluxes in type IIA deform the topology type of the internal space, making the K\"ahler and/or complex structure forms non-closed~\cite{Frey:2003tf}. In particular, the flux superpotential receives corrections $W \supset \int dJ \wedge \Omega$, with $||dJ|| \sim \sqrt{\vol}/L\sim N^{-3/2}$; these are negligibly small at large $N$. More generally, the backreaction induces a nonzero Ricci scalar $\mc R^{(6)} \sim 1/L^2$, modifying the effective potential by an amount
$$
V \sim \left(\frac{g_s^2}{\vol} \right)^2 \frac{\vol}{g_s^2}\,\mc R^{(6)} \sim \frac{1}{N^3}\,\frac{1}{L^2}\,,
$$
where the first factor comes from the Einstein frame. Such corrections are parametrically small compared to the terms that we have kept in the potential.

An important additional contribution comes from the nonzero chemical potentials. The EFT should remain valid as long as these are much smaller than the KK scale. To illustrate this effect, consider corrections to the moduli kinetic terms and gauge kinetic function depending on some scalar KK mode $\phi_{KK}$. By orthogonality of the internal wavefunctions, the lowest possible such correction is of order $\phi_{KK}^2$. Then, as long as the order of magnitude of the black brane energy is much smaller than $m_{KK}$, the corrections on the $\phi_{KK}$ equation of motion (and hence on the moduli effective action) will be negligible. The precise form of such effects depends on each CY, but the main point is that
because of the parametric separation of scales, KK modes can't be destabilized.

We should stress that there are nonperturbative effects, not included in the EFT, from Schwinger pair-production of the charge carriers and of color branes that give rise to the dual gauge groups. In the gravity side these are modeled, respectively, by D2s wrapped on the 2-cycles that support the gauge fields, or Dp branes that are domain walls in the 4 external directions. The motion of these probe branes serves as tests for possible nonperturbative instabilities in the dual QFT, like the ``Fermi sea-sickness'' phenomenon of~\cite{Hartnoll:2009ns}. An analysis of such effects is left to future work~\cite{future}.

\section{Black brane solutions}\label{sec:solutions}

In the remaining of the paper we will analyze explicit black brane solutions in flux compactifications. The range of possible phases in these theories appears to be extremely rich, and here we will only take the first steps towards understanding them. The simplest exact solutions are $AdS_2 \times \mathbb R^2$, which are interesting for dual descriptions of emergent local quantum criticality. Next we turn to generalizations of the Reissner-Nordstrom AdS (RN-AdS) black hole, where the cosmological constant arises from expectation values of scalars that evolve radially. Finally, we discuss branes that exhibit hyperscaling violation. For concreteness, these solutions are studied in the simple models of \S \ref{sec:models}. It would be very interesting to develop tools to analyze black branes at the level of the general EFT (some of which are described in \S \ref{subsec:hyperscaling}), and understand the landscape of finite density flux vacua.

\subsection{$AdS_2\times \mathbb R^2$ fixed point}\label{subsec:AdS2}

The simplest solution is an extremal $AdS_2\times \mathbb R^2$ geometry. This geometry has been studied as
the zero temperature limit of the RN-AdS black hole, in order to describe systems with emergent local quantum criticality~\cite{Liu:2009dm}. It also appears as an exact solution in our setup. The extremal solution is characterized by having a double pole for the emblackening factor $f(r) \sim (r-r_h)^2$ at the horizon. 

The exact solution is found in a new radial coordinate $u$, in terms of which the $AdS_2\times \mathbb R^2$ metric takes the form
\be\label{eq:new-ansatz}
ds^2 = - \frac{u^2}{L_{(2)}^2} dt^2 + \frac{L_{(2)}^2}{u^2} du^2 + b_h^2 (dx_1^2 + dx_2^2)\,.
\ee
Here $L_{(2)}$ is the $AdS_2$ radius and $b_h^2 = r_h^2/L^2$ in the ansatz (\ref{eq:metric-ansatz}). The equations of motion from the radial Lagrangian (\ref{eq:radialL}) in this new coordinate system then imply that the scalars are stabilized at
\be
\frac{\partial}{\partial \phi^I}\,\log \Vf(\phi_h)= \frac{\partial}{\partial \phi^I}\,\log \Vb(\phi_h)
\ee
and the $AdS_2$ radius and horizon position are given by
\be
L_{(2)}^2 = - \frac{1}{2 \Vf(\phi_h)}\;,\;b_h^4 = \frac{r_h^4}{L^4}=\frac{\Vb(\phi_h)}{- \Vf(\phi_h)}\,.
\ee
See also~\cite{Goldstein:2009cv}. An intriguing property of these solutions, noted many times in the past, is their ground state entropy density
\be\label{eq:AdS2entropy}
 \frac{\mc S}{\mV}= \frac{2\pi}{\kappa_4^2} \sqrt{\frac{\Vb(\phi_h)}{- \Vf(\phi_h)}}\,.
\ee

Let us obtain the $AdS_2\times \mathbb R^2$ in some of the simple models. In the nonsupersymmetric toy model of
\S \ref{subsec:simplified}, 
$$
\Vf=-Ae^{-\alpha\nu_1\phi}+Be^{-\alpha\nu_2\phi},\;\Vb=\frac{1}{2}(Q^2e^{-\alpha\phi}+P^2e^{\alpha\phi})\,.
$$
A few algebraic manipulations reveal that the horizon value of the scalar field satisfies
\be\label{eq:dyonic}
\frac{\nu_2-\nu_1}{2}\coth{(\alpha\frac{\nu_2-\nu_1}{2}\phi_h+\delta_2)}=\tanh{(\alpha\phi_h+\delta_2)}+\frac{\nu_2+\nu_1}{2}
\ee
and the size of the horizon is
\be
\frac{r_h^4}{L^4}=\frac{PQ\cosh{(\alpha\phi_h+\delta_2)}}{\sqrt{AB}\sinh{(\alpha\frac{\nu_2-\nu_1}{2}\phi_h+\delta_1)}}e^{\alpha\frac{\nu_1+\nu_2}{2}\phi_h}
\ee
where $\delta_1=\sinh^{-1}{(\frac{A-B}{2\sqrt{AB}})},\;\delta_2=\sinh^{-1}{(\frac{P^2-Q^2}{2PQ})}$. Given that $\nu_1,\nu_2>0,\nu_2>\nu_1$, one can show that (\ref{eq:dyonic}) always admits a solution of $\phi_h$ such that $r_h^4>0$. We hence conclude that there always exists dyonic near horizon $AdS_2\times \mathbb R_2$ solutions in this model.  For a purely electrically charged brane, the zero temperature near horizon solution can be given explicitly,
\be\label{eq:T0}
\alpha \phi_h=\frac{1}{\nu_2-\nu_1}\log\left(\frac{\nu_2-1}{\nu_1-1}\frac{B}{A}\right)\,.
\ee
This near horizon value is independent of the electric charge, which cancels because it is the only dimensionful perturbation. In contrast, the dyonic solution (\ref{eq:dyonic}) depends on the dimensionless coupling $\delta_2$.

Now we also study the theory \S\ref{subsec:oneK} with a single K\"ahler-modulus and look for an electrically charged $AdS_2\times \mathbb R^2$ solution. Absent magnetic charges, the axions are stabilized at their supersymmetric values, so we only need to consider the K\"ahler modulus $v=e^{\phi}$ and dilaton $D$. The near horizon values are found to be
\be\label{eq:AdS_2 vev}
v_h=(6+4\sqrt{3})^{\frac{1}{4}}\sqrt{-\frac{\t e_1}{\kappa m_0}}\;,\;
e^{-D_h}=(-6+4\sqrt{3})^{\frac{1}{4}}\frac{\sqrt{3}}{2}\,\frac{\kappa m_0}{p} \left(-\frac{\t e_1}{\kappa m_0} \right)^{3/2}\,,
\ee
which have the same parametric dependence on the fluxes as the supersymmetric vacua (\ref{eq:AdS4vev}). Recall that $\t e_1 \sim N \gg 1$ in the large flux limit where we have perturbative control.
As before, the dependence on the electric charge cancels for dimensional reasons.
The ground state entropy density is given by
\be
\frac{\mc S}{\mc V} = \frac{2\pi}{\kappa_4^2}\sqrt{\frac{1}{2}+\frac{1}{\sqrt{3}}}\,\frac{4}{3}\, \frac{\t e_1^2Q}{\sqrt{\kappa m_0^2 p^4}} \sim N^2 Q\,.
\ee

The dimensions of operators in the dual quantum critical point can be obtained by expanding (\ref{eq:oneK}) plus the contribution $\Vb$ from the charge density around the new horizon values for the scalars. The physical masses in $AdS_2$ units are
\be\label{eq:massAdS2}
M^2_\text{k}L_{(2)}^2=(12.10,2.41)\;,\;\;M^2_\text{axion}L_{(2)}^2=(13.9,1.51)\,.
\ee
In the next section we argue that there exists a solution that interpolates between $AdS_4$ in the UV and this IR geometry.
In the dual theory at finite chemical potential, this describes an RG flow between a 3d UV fixed point and a conformal quantum mechanics in the IR (or perhaps the chiral sector of a 2d CFT). The only scalar operators of small dimension in the UV theory have $\Delta =5,\,6,\,10,\,11$ [see (\ref{eq:massK})], which flow to $\Delta =1.82,\,2.13,\,4.01,\,4.26$ in the IR. It would also be interesting to study the fermionic sector~\cite{future}.

The existence of $AdS_2\times \mathbb R^2$ vacua in black brane solutions of flux compactifications provides a UV complete setup to study holographic systems with emergent local quantum criticality. It would be especially interesting to understand the microscopic origin of the ground state entropy (\ref{eq:AdS2entropy}) and their stability. As illustrated in the previous example, it is possible to construct dual theories with no relevant operators which, at finite density, are free of homogeneous instabilities at the perturbative level. To establish this, it is important that KK modes decouple. There could also be inhomogeneous instabilities, like the one found in~\cite{Donos:2011bh}, but our analysis so far shows that there can be stable solutions~\cite{future}.

\subsection{Generalized Reissner-Nordstrom branes}\label{subsec:RN}

A distinguishing property of black branes in flux compactifications is that the cosmological constant arises as the expectation value of the potential.  A `hard' cosmological constant gives the RN-AdS geometry, but now the nontrivial radial evolution of the scalars will introduce corrections. In general it is then hard to find analytic solutions. 
The procedure to construct black branes starts from perturbative expansions near the UV and IR regions, and then finds the full interpolating geometry numerically. The perturbation series have some undetermined integration constants, which are fixed in the full solution. One approach to this problem is to use the perturbative expansion near the horizon to seed the numerical integration, and then scan over the integration constants until the correct UV asymptotics is reproduced. 

In the asymptotically $AdS_4$ region ($r \gg L$ in our coordinate system), the normalizable modes are the expectation values for the scalar fields, and we fix their sources to zero. We deform from the $AdS_4$ solution by turning on a metric mode $f(r) \propto f_1 L/r$, as well as nonvanishing temporal and spatial components for the bulk gauge fields. These correspond to nonzero temperature (\ref{eq:temperature}), chemical potentials (\ref{eq:chemical potential}) and magnetic fields (\ref{eq:magneticB}).

Given these sources, the solution of (\ref{eq:radialL}) at large $r$ is
\bea\label{eq:UVseries}
f(r) & = & \frac{r^2}{L^2} - f_1 \frac{L}{r} + L^2 \langle \Vb \rangle \frac{L^2}{r^2} +f^I \frac{L^{2(\Delta_I-1)}}{r^{2(\Delta_I-1)}}+ \ldots \nonumber\\
\phi^I(r) &=& \langle \phi^I \rangle + c^{IJ} \frac{L^{\Delta_J}}{r^{\Delta_J}}+L^2d^I\frac{L^4}{r^4}+L^2n^I\frac{L^7}{r^7} + \ldots \\
w(r) &=& 4L^4K^{I\bar{J}}d_Id_{\bar{J}}\frac{L^8}{r^8} + 2 K_{I \bar J} c^{IL} \bar c^{JN} \frac{\Delta_L \Delta_N}{\Delta_L + \Delta_N} \frac{L^{\Delta_L + \Delta_N}}{r^{\Delta_L + \Delta_N}} + \ldots \nonumber
\eea
where $d^I$, $f^I$ and $n^I$ depend on the flux and brane potentials, and the dimensions $\Delta$ are given by (\ref{eq:Delta}). Summation over repeated indices is implicit,
with different fall-offs appearing because in general the $\phi^I$ are not mass eigenvectors.\footnote{The series for $\phi^I$ is schematic because, as we explained before, the scalar and pseudoscalar parts of $\phi$ have different masses, so one should write the two series separately.} Also, $\langle \ldots \rangle$ denote expectation values in the $AdS_4$ vacuum. Eqs.~(\ref{eq:UVseries}) show the leading effects from the temperature and electric and magnetic sources. The first three terms in $f(r)$ give the RN-AdS black brane, and the corrections to this behavior arise because the scalar fields are dynamical. At this stage, the only arbitrary integration constants are the $c^{IJ}$, which are dual to expectation values of single trace operators $\mc O^I$ in the gauge theory, and $f_1$ that will be related to the temperature below.

Let us now turn our attention to the IR region. At the horizon $r=r_h$, the warp factor $f(r) \sim r - r_h$. The solution is required to be regular at the horizon, so we expand
\bea\label{eq:IRseries}
f(r) &=& - r_h^2 V_\text{eff}(r_h,\phi_h)\, \frac{r-r_h}{r_h} \ldots \nonumber\\
\phi^I(r) &=& \phi_h^I - K^{I \bar J} \partial_{\bar J} V_\text{eff}(r_h,\phi_h)\,\frac{r-r_h}{r_h}  + \ldots\\
w(r) & = & w_h -2\,K^{I \bar J} (\partial_{I} \log V_\text{eff}(r_h,\phi_h))(\partial_{\bar J} \log V_\text{eff}(r_h,\phi_h))\, \frac{r-r_h}{r_h}  + \ldots \nonumber
\eea
where the effective radial potential (here evaluated at the horizon) was introduced in (\ref{eq:Veff}).
From this near horizon expansion, the arbitrary integration constants are $r_h$, $\phi_h^I$ and $w_h$. 

The full black brane solution can be obtained by a standard shooting method, where the IR series (\ref{eq:IRseries}) (to high enough order) is used to seed the numerical integration very close to the horizon. The arbitrary constants $w_h$ and $\phi_h^I$ are adjusted until the correct UV asymptotics (\ref{eq:UVseries}) is reproduced. This procedure fixes $(c^{IJ},f_1,\phi_h^I,w_h)$ in terms of the parameters $(r_h, \mu^\alpha, B^\alpha)$. The expectation values of the dual single trace operators (extracted from the $c^{IJ}$) are in general nonzero.

Let us illustrate this for the non-supersymmetric model (\ref{subsec:simplified}) --the procedure for other models with more scalars is similar but longer to present. As an example, consider $\nu_1=2$, $\nu_2=3$, $\alpha=1$, and $A=\frac{3\nu_2}{\nu_2-\nu_1},B=\frac{3\nu_1}{\nu_2-\nu_1}$, so that $\langle\phi\rangle=0$ and $L=1$.  This describes a holographic $(2+1)$--dimensional CFT with a single scalar operator of dimension $\Delta=6$ and one global symmetry. We study electric branes, corresponding to putting this CFT at finite charge density.
As discussed before, we start from the near horizon solution, and ``shoot'' for the correct $AdS_4$ asymptotics. The series expansion depends on two parameters $\phi_h$ and $w_h$. The equation of motion for $w(r)$ contains one additive integration constant, which is determined by a boundary condition either at the horizon or the boundary. The usual choice is
$w(\infty)=0$, but here we shall simply set $w_h=0$ instead. The two boundary conditions simply differ by a rescaling of $t$. Then we are left with only one parameter $\phi_h$. The spacetime coordinates are rescaled to set $r_h=1$ and the black brane is then uniquely specified by the dimensionless charge density $Q$ in units of temperature.
The temperature and chemical potential can be extracted from the numerical solution, via (\ref{eq:temperature}) and (\ref{eq:chemical potential}). 

\begin{figure}[htb]
\begin{center}
\includegraphics[scale=0.89]{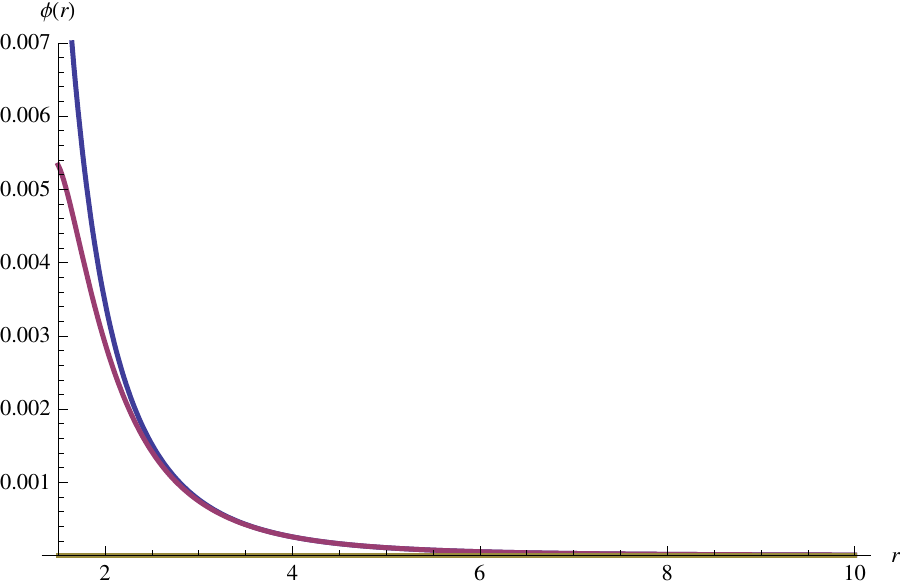}
\includegraphics[scale=0.89]{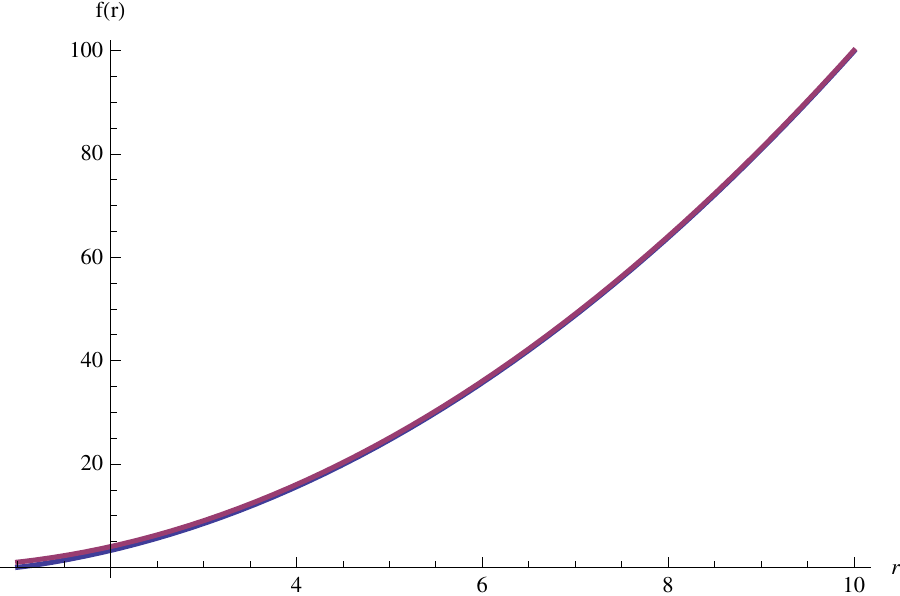}
\includegraphics[scale=0.89]{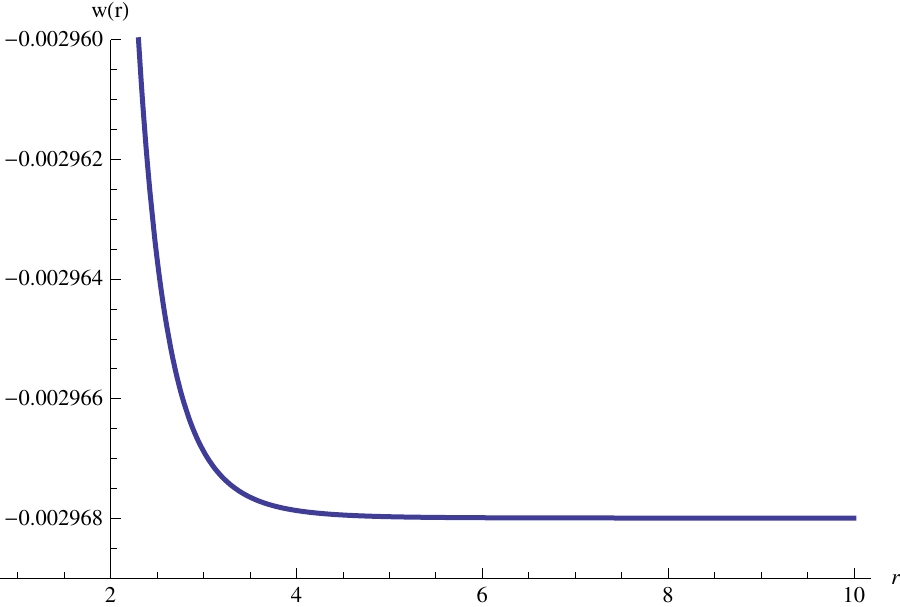}
\includegraphics[scale=0.89]{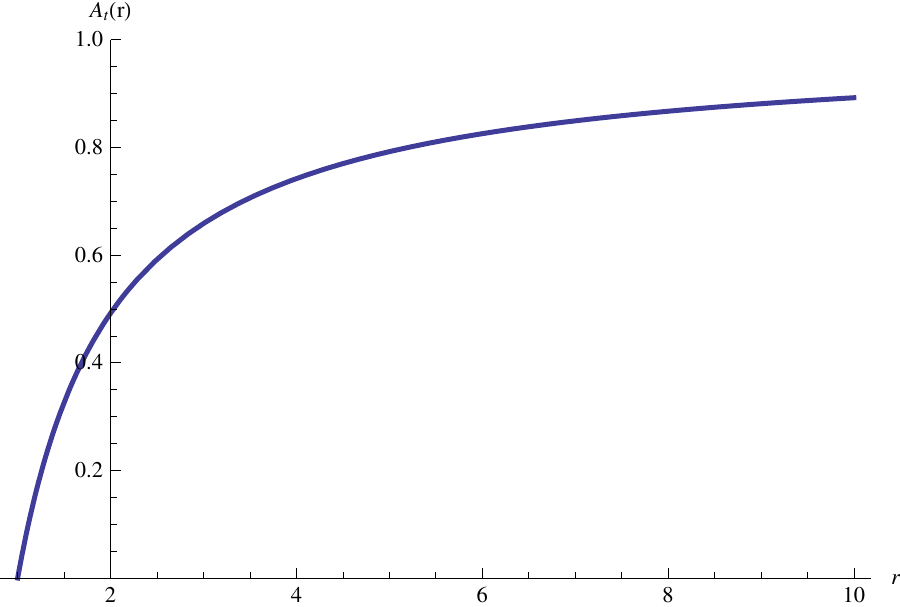}
\caption{{\small Top left figure plots $\phi(r)$(blue) and the UV series expansion $\frac{Q^2}{28}\frac{1}{r^4}-\frac{0.1}{r^6}$(purple); top right figure plots $f(r)$(blue) and the asymptotic $AdS_4$ behavior; bottom left and bottom right are plots of $w(r)$ and $A_t(r)$ respectively, from which we can extract the temperature and chemical potential.}}
\label{fig:plot2}
\end{center}
\end{figure}

\begin{figure}[t]
\begin{center}
\includegraphics[width=0.6\textwidth]{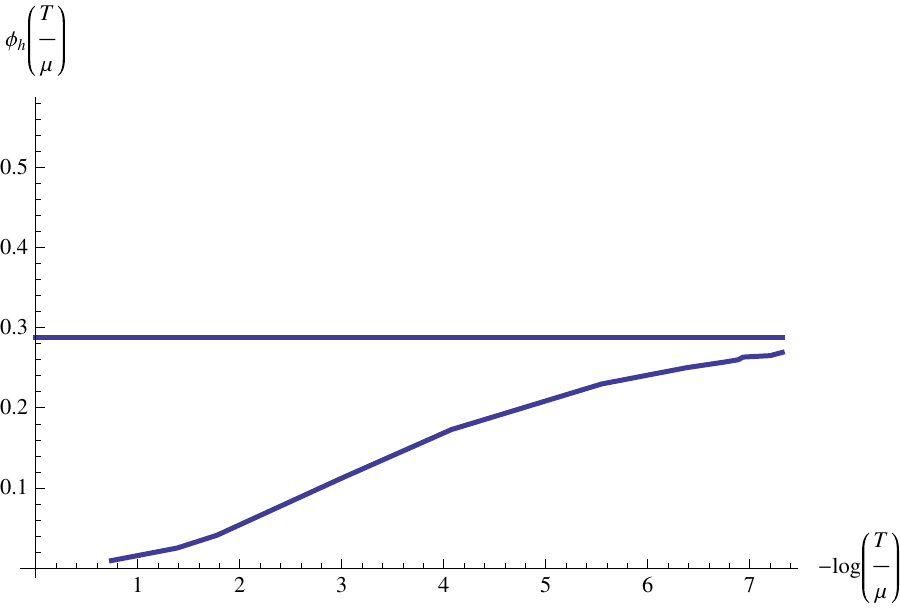}
\caption{{\small Plot of $\phi_h(\frac{T}{\mu})$ against the dimensionless scale $-\log{\frac{T}{\mu}}$, compared with the zero temperature value $\phi_h(T=0)=\log{\frac{4}{3}}$.}}
\label{fig:plot3}
\end{center}
\end{figure}

For each $Q$, we scan over the horizon values $\phi_h$ until the scalar field profile has the correct asymptotic behavior, $$
\phi(r)\sim \frac{Q^2}{28r^4}+\frac{c}{r^{\Delta}}
$$
as $r\to \infty$, with $\Delta=6$. The coefficient $c$ is the expectation value of the dual operator at finite $T$ and $\mu$. It can then be checked that the metric approaches the correct $AdS_4$ geometry when this happens. Figure \ref{fig:plot2} shows the numeric solution for $Q= \sqrt{2}$, for which the shooting method gives the horizon value $\phi_h = 0.0410$ and expectation value $c \sim -0.1$ for the dual operator. From the full numerical solution, we can recover the standard $\tilde{w}(r)\sim w(r)-w(\infty)$, hence $\tilde{w}_h\sim -w(\infty)=0.0297$. The temperature and chemical potential of the brane can then be extracted by
\be
T=\frac{1}{4\pi}f'(r_h)e^{-\frac{\tilde{w}_h}{2}}=1.61,\;\mu=A_t(\infty)=\int_{r_h}^{\infty}dr e^{-\frac{w}{2}-\phi}\frac{L^2Q^2}{2r^2}=0.892\nonumber
\ee
 We also examine how the solution varies with charge density $Q$, and verify that the $AdS_2 \times \mathbb R^2$ solution arises as the zero temperature limit of the finite temperature geometry.
Figure \ref{fig:plot3} plots the horizon values $\phi_h$ as a function of $\log{\frac{T}{\mu}}$ and the limiting $AdS_2 \times \mathbb R^2$ attractor value $\phi_h=\log{\frac{4}{3}}$. This strongly suggests that the IR $AdS_2\times \mathbb R^2$ fixed point can be connected to the $AdS_4$ vacuum.

\subsection{Branes with hyperscaling violation}\label{subsec:hyperscaling}

Finally, we study solutions with hyperscaling violation $\theta$ and dynamical exponent $z$,
\be\label{eq:hyperansatz}
f(r) = f_0 \,r^{4/(2-\theta)}\;,\;e^{-w(r)} = r^\omega\,,
\ee
where we have introduced the shorthand notation
\be\label{eq:omegadef}
\omega \equiv 2 \,\frac{2z-2-\theta}{2-\theta}\,.
\ee
These geometries transform covariantly under scalings, and describe holographic field theories whose thermodynamics behaves as if they lived in $2-\theta$ effective spatial dimensions --as can be seen from the thermal entropy $\mathcal S \sim T^{(2-\theta)/z}$. The case $\theta=1$ is a promising candidate for Fermi surfaces~\cite{Ogawa:2011bz}, and more general values of $\theta$ can reproduce some of the properties of systems with disorder~\cite{disorder}. In UV complete theories, these geometries are in general valid over some finite range of $r$, with simple examples coming from $Dp$ brane solutions~\cite{Dong:2012se,Dey:2012tg}.

Let us first discuss some general properties. The range of allowed $z$ and $\theta$ is constrained by the null energy condition~\cite{Dong:2012se},
\be\label{eq:nec}
(2-\theta) (2z-2-\theta) \ge 0\;,\;(z-1) (2+z-\theta) \ge 0\,.
\ee
Note that this implies $\omega \ge 0$ in (\ref{eq:omegadef}). Furthermore, we require the existence of a finite temperature solution valid for $r > r_h$, and we have assumed that the extremal horizon is at $r=0$.\footnote{This requires $g_{tt} \to 0$ as $r \to 0$. The conclusion below regarding the source of negative energy is also valid if the singularity is at large $r$, but a holographic interpretation is not clear in this case.} This implies
\be\label{eq:naked}
\frac{2}{\theta-z}\le 1\,.
\ee
Next, replacing the hyperscaling ansatz (\ref{eq:hyperansatz}) in the radial equations (\ref{eq:radialeqs}) obtains
\be\label{eq:whyperscaling}
K_{I \bar J} \partial_r \phi^I \partial_r \bar \phi^{\bar J} = \frac{\omega}{2r^2}
\ee
and
\be\label{eq:fhyperscaling}
f_0 r^{\frac{4}{2-\theta}} \left(2+ \frac{2z}{2-\theta} \right)=-r^2 \left(\Vf(\phi) + \frac{L^4}{r^4} \Vb(\phi) \right)\,.
\ee

The conditions (\ref{eq:nec}) and (\ref{eq:naked}) imply that the left hand side of this last equation is positive. We thus reach the general conclusion that hyperscaling violating geometries require a \textit{source of negative energy} that has to dominate the total potential energy. We stress that this is a local statement, independent of whether such geometries can be embedded into an asymptotically AdS solution. Such a source can come from positive internal curvature, for instance in sphere reductions~\cite{Perlmutter:2012he} or near horizon limits of branes~\cite{Dong:2012se,Dey:2012tg}. For the flux compactifications discussed in this work, the source of negative energy comes from orientifold planes.

To gain intuition, it is useful to first analyze hyperscaling violating solutions in the nonsupersymmetric model. For a purely electric solution, balancing the negative term in the potential (\ref{eq:Ssimplified}) (which we just argued has to dominate in the hyperscaling violating regime) against $\Vb$ gives a solution with
\be
\theta = \frac{4 \nu_1}{\nu_1+1}\;,\;z= 3- \frac{2}{\nu_1+1} - \frac{8}{\alpha^2(\nu_1^2-1)}
\ee
and $\phi \sim \frac{4}{\alpha(\nu_1-1)} \,\log r$. The solution is valid approximately at large $r$ if $\nu_1>1$, and at small $r$ otherwise. These results agree with the electric branes discussed in~\cite{Iizuka:2011hg}. On the other hand, if the magnetic charge is nonzero, it always dominates over the electric charge, yielding parameters
\be
\theta = \frac{4\nu_1}{\nu_1-1}\;,\;z= 3+ \frac{2}{\nu_1-1} - \frac{8}{\alpha^2(\nu_1^2-1)}
\ee
with $\phi \sim \frac{4}{\alpha(\nu_1+1)} \,\log r$. The solution is valid at large $r$. This agrees with the magnetic branes of~\cite{Bhattacharya:2012zu}. A hyperscaling violating solution has positive specific heat if
\be\label{eq:thermal-stab}
\frac{2-\theta}{z} \ge 0\,,
\ee
which in this case translates into
\be
0 \le \nu_1 \le \pm \frac{1}{3} + \frac{2}{3} \sqrt{1+6/\alpha^2}
\ee
for the electric/magnetic branes respectively. 

Let us now consider hyperscaling violation in CY flux compactifications. From (\ref{eq:fhyperscaling}), the flux and brane potential energy contributions
\be
\Vf \sim r^{2\theta/(2-\theta)}\;,\;\Vb \sim r^{(2\theta-8)/(\theta-2)}\,.
\ee
The orientifold is the only source of negative energy and it has to dominate the total energy. This determines the behavior of the physical dilaton
\be
D = D_0 + \frac{2}{3} \frac{\theta}{2-\theta}\,\log r\,.
\ee
Eq.~(\ref{eq:whyperscaling}) implies that the K\"ahler moduli $v_i$ have power-law dependence on $r$, and finding a hyperscaling violating solution requires identifying consistently a set of terms in $\Vf$ and $\Vb$ that give the dominant contribution.

In practice, obtaining these solutions analytically is complicated by the fact that these compactifications generically have many scalars. However, in the single K\"ahler modulus theory of \S \ref{subsec:oneK} this can still be done by direct inspection. We find a consistent electric solution where axions do not run,
\be
v \sim r^{1/2}\;,\;D \sim - \frac{3}{2}\,\log r
\ee
and
\be
\theta = \frac{18}{5}\;,\;z=\frac{17}{20}\,.
\ee
The solution is valid at large $r$ and is under perturbative control because $\vol \to \infty$ and $g_s \to 0$. It is supported by a combination of electric charge, orientifold tension, $H_3$ flux and Romans mass. Nevertheless, $\theta>2$, so the specific heat is negative and there may be instabilities. It would be interesting to study the stability of this solution, along the lines of~\cite{Cremonini:2012ir}. This motivates the more general question of a possible ``correlated stability conjecture''~\cite{Gregory:1993vy}  for hyperscaling violation geometries, which would be worth analyzing.

\subsubsection{General racetrack potentials}

It is of interest to have more efficient tools to study hyperscaling violating solutions in the flux landscape. Here we take the first steps in this direction by obtaining some general results for racetrack potentials $V = \sum A_m \exp(\sum \alpha_{m,i} \phi_i)$.

We assume that in a hyperscaling violating geometry (\ref{eq:hyperansatz}), the canonically normalized scalar fields are running in the form 
\be
\phi_i(r)=\log \phi_i^0+k_i\log r\,,
\ee
which is the case for kinetic terms in perturbative flux compactifications.
The effective potential is then a polynomial in $r$. Usually a hyperscaling violating geometry is an approximate solution, sourced by a subset of terms in $V_\text{eff}$ that have to consistently dominate in the regime of interest. Now we are going to study how the hyperscaling violating exponents are constrained in our EFT.

The leading order equations of motion are
\bea
f_0^2\frac{\theta-6+(\frac{\theta}{2}-1)\sum k_i^2}{\theta-2}r^{\frac{2\theta}{2-\theta}}+V_\text{eff}=0\\
\frac{4+2\theta-4z}{2-\theta}+\sum k_i^2=0\label{eq:beta}\\
f_0^2\frac{\theta-6+(\frac{\theta}{2}-1)\sum k_i^2}{\theta-2}k_m r^{\frac{2\theta}{2-\theta}}-\partial_{\phi_m} V_\text{eff}=0
\eea 
Assume that the leading order terms in the potential comes from the following subset of terms from the full potential,
\be
V_\text{dominate}=\frac{V_\text{brane}}{r^4}+\sum_m A_m\exp{\sum \alpha_{m,i}\phi_i}\,.
\ee
For simplicity, we consider a single gauge field and denote by $\phi_1$ the scalar field in its gauge kinetic function; then $\Vb =\frac{1}{2} Q^2 e^{\pm \alpha \phi_1}$, with the sign determined by whether the brane is electric or magnetic. For $V_\text{dominate}$ to source the geometry to the leading order, it needs
\be\label{eq:power}
\sum_i \alpha_{i,m} k_i=\frac{2\theta}{2-\theta}\;\;,\;\;
\pm \alpha k_1-4=\frac{2\theta}{2-\theta}\,.
\ee

It is useful to introduce the notation
\be
\tilde{A}_m \equiv A_m \prod (\phi_i^0)^{\alpha_{i,m}}\;\;,\;\;\tilde{A}_0 \equiv \frac{Q^2}{2}(\phi_1^0)^{\alpha_{0,1}}\;,\;\alpha_{0,i}\equiv \pm \alpha \delta_{1,i}
\ee
which make the leading order equations of motion algebraic,
\bea
f_0^2\,\frac{\theta-6+(\frac{\theta}{2}-1)\sum k_i^2}{\theta-2}+\sum_n \tilde{A}_n=0\label{eq:a0}\\
f_0^2\,\frac{\theta-6+(\frac{\theta}{2}-1)\sum k_i^2}{\theta-2}\,k_j-\sum_n\tilde{A}_n \alpha_{j,n}=0
\eea
We further define an auxiliary ``probability'' 
\be
P_l \equiv \frac{\tilde{A}_l}{\sum_n \tilde{A}_n}\;\;,\;\;\langle f \rangle \equiv \sum_n f_n P_n
\ee 
(a slight abuse of name, since $P_l$ is not necessarily positive), and solve for $k_i$
\be
k_i=-\ai = - \sum_n \alpha_{in} P_n\,.
\ee
One can then show that the hyperscaling violation and dynamical exponents are
\be
\theta=2+\frac{4}{\sum_i \ai^2+4P_0-2},\;z=1+\frac{4P_0}{\sum_i \ai^2+4P_0-2}\,,
\ee
and
\be\label{eq:positivity}
f_0^2=\frac{1}{4P_0+\frac{1}{2}\sum_i \ai^2-3}\sum_n \tilde{A}_n\,.
\ee

The exponents $(\theta, z)$ are determined in terms of the auxiliary probabilities $\{P_l\}$, which are in turn fixed by
the power-matching identities (\ref{eq:power}). From the definition of $\tilde{A}_n$, $P_l$ are functions of $\{\phi_i^0\}$,  hence if the number of scalars is greater than the number of terms in $V_\text{dominate}$, it is in general possible to solve for $\{\phi_i^0\}$'s given the $\{P_l\}$. However, additional constraints need to be satisfied in order to have a physically sensible solution. In particular, one should check  that $V_\text{dominate}$ self-consistently dominates over the neglected terms in $V_\text{eff}$, and that $\phi^0_i$ are real and $f_0 >0$. Furthermore, the condition that the total potential energy in the hyperscaling violating regime be negative now translates into $\sum_n\tilde{A}_n<0$.
We may also want to impose positivity of the specific heat, which amounts to
\be
1-4P_0-\frac{1}{2}\sum_i \ai^2\ge0\,,
\ee  
although, as we discussed before, there may be interesting phases arising from instabilities.

\section{Summary and future directions}\label{sec:concl}

In the present work we have constructed charged black branes in flux compactifications of type IIA string theory on CY manifolds. The gauge fields arise from the 3-form RR potential evaluated on harmonic 2-forms and the essential feature of these solutions is that the six internal dimensions are parametrically smaller than all the relevant scales. Black branes are described in terms of a 4d effective field theory that includes only a few light fields and where KK modes decouple. In the perturbative regime of large flux, the EFT depends only on topological information of the CYs and not on specific metric properties. We also studied basic aspects of the $(2+1)$--dimensional duals at finite chemical potential, including their spectrum of operators and thermodynamic properties.

Clearly a lot of work remains to be done to understand the holographic phases of matter described by flux compactifications at finite density. We focused on some of the simple solutions, including $AdS_2 \times \mathbb R^2$, generalized Reissner-Nordstrom and branes with hyperscaling violation. These systems may exhibit phenomena that are qualitatively different from known Freund-Rubin compactifications. 
Along this direction, it would be important to study transport properties of the charge carriers, which are dual to operators of very high dimension (unlike examples where the charges descend from internal isometries).

The appearance of emergent local quantum criticality in the flux landscape is very intriguing, and the stability and holographic properties of such solutions are currently under investigation~\cite{future}. We found hyperscaling violating solutions with negative specific heat. It would be interesting if no-go theorems can be found for these theories, and also if an analog of the correlated stability conjecture~\cite{Gregory:1993vy} exists for hyperscaling violation. Given the simple scaling properties of such geometries, these questions may be tractable. Other phases that could potentially be UV-completed in flux compactifications include the multi-centered solutions of~\cite{Anninos:2011vn}, and homogeneous but anisotropic branes~\cite{Iizuka:2012iv}.

Finally, it would also be interesting to explore different limits of string vacua. For instance, strong warping leads to nontrivial kinetic terms~\cite{Douglas:2007tu}, and one could consider the effects of finite density in simple geometries of this type. Also,~\cite{Polchinski:2009ch} found $AdS_5$ solutions with small extra dimensions using $(p,q)$ 7-branes, and it would be interesting to analyze such 4d CFTs at finite chemical potential. Finally, type IIB AdS solutions on CYs with nonperturbative effects~\cite{Kachru:2003aw} provide another framework for charged brane solutions that is worth analyzing.

\section*{Acknowledgments}

We are particularly grateful to S.~Kachru for collaboration during the initial stages of this project, and for constant encouragement and interesting discussions. We have benefited from helpful discussions with S.~Hartnoll, E.~Silverstein and T.~Wrase.
We would also like to thank D.~Anninos, N.~Bao, S.~Cremonini, S.~Harrison, S.~Hartnoll, S.~Kachru, and S.~Trivedi for comments on a draft of this work.
The research of G.T.~is supported in part by the National Science Foundation under grant no.~PHY-0756174. H.W.~is supported by a Stanford Graduate Fellowship.

\appendix

\bibliographystyle{JHEP}
\renewcommand{\refname}{Bibliography}
\addcontentsline{toc}{section}{Bibliography}
\providecommand{\href}[2]{#2}\begingroup\raggedright

\end{document}